\documentclass[aps,prb,reprint,superscriptaddress,floatfix]{revtex4-2}

\usepackage{amsmath,amssymb,bm}
\usepackage{graphicx}
\usepackage[dvipsnames]{xcolor}
\usepackage{hyperref}
\hypersetup{colorlinks=true,linkcolor=blue,citecolor=blue,urlcolor=blue}

\graphicspath{{./}}

\newcommand{\GET}{\Gamma_{\mathrm{ET}}}
\newcommand{\lX}{\ell_X}
\newcommand{\muF}{\mu_F}
\newcommand{\qc}{q_c}
\newcommand{\vF}{v_F}

\begin{document}

\title{Graphene as a Tunable Nonradiative Bath for Moir\'e Excitons}

\author{Katsunori Wakabayashi}
\email{WAKABAYASHI.Katsunori@nims.go.jp}
\affiliation{Research Center for Materials Nanoarchitectonics (MANA),
National Institute for Materials Science (NIMS),
Namiki 1-1, Tsukuba 305-0044, Japan}

\date{\today}

\begin{abstract}
A minimal theory of nonradiative energy transfer from a
two-dimensional (2D) moir\'e exciton to a nearby graphene layer is presented. From Fermi's
golden rule the transfer rate is the overlap of the exciton near-field spectrum
with the dissipative density response of graphene, weighted by an exciton form
factor, and it reproduces the established $\GET\propto z^{-4}$ law in the
point-dipole limit. A finite exciton size filters out the high-momentum near field
once the spacer thickness approaches the transition-polarization radius $R_X$, so
the distance dependence of the rate---and of the photoluminescence (PL)
quenching---probes the exciton size. A low-momentum expansion shows that, relative
to the calibrated point-dipole response of the same bath, this leading correction
is set by $R_X$ alone. In the ideal coherent-envelope limit the accompanying giant
oscillator strength makes the rate non-monotonic in the exciton size, with a peak
near $\lX\approx z$. Treating graphene as a gate-tunable bath, Pauli blocking
suppresses the interband channel once $2|\muF|$ approaches $\hbar\omega$, partially
restoring PL, and a full random-phase-approximation benchmark confirms the
normalized interband distance dependence to within a few percent away from the
threshold. Mapping the PL observables across the transition-metal
dichalcogenide/hexagonal boron nitride/graphene parameter space, we find that a
graphene gate acts not as a passive electrostatic element but as a tunable 2D
electronic reservoir probed through exciton PL quenching.
\end{abstract}

\maketitle

\section{Introduction}

When an excited emitter is placed near a conducting or semimetallic system, it
can relax by transferring its energy nonradiatively to electronic excitations of
that system instead of emitting a photon. For graphene, Swathi and Sebastian
showed using a tight-binding model and the Dirac-cone approximation that the
resonance energy-transfer (ET) rate from a point dye molecule scales as
$z^{-4}$ with the emitter--graphene distance $z$~\cite{Swathi2008,Swathi2009},
much longer ranged than the $R^{-6}$ F\"orster law of molecular
fluorescence resonance energy transfer~\cite{Forster1948}. This distance scaling
has been observed experimentally for individual quantum emitters and
semiconductor nanostructures~\cite{Gaudreau2013,Chen2010,Federspiel2015}, and the relaxation
pathway has been shown to be electrically controllable through the graphene Fermi
level~\cite{Tielrooij2015,Lee2014}. On the theory side, the graphene fluorescence-quenching
problem has been revisited with the full electronic response of graphene,
clarifying the role of its loss function and of longitudinal and transverse
plasmons~\cite{GomezSantos2011,Velizhanin2011}.

\begin{figure*}[t]
  \centering
  \includegraphics[width=\textwidth]{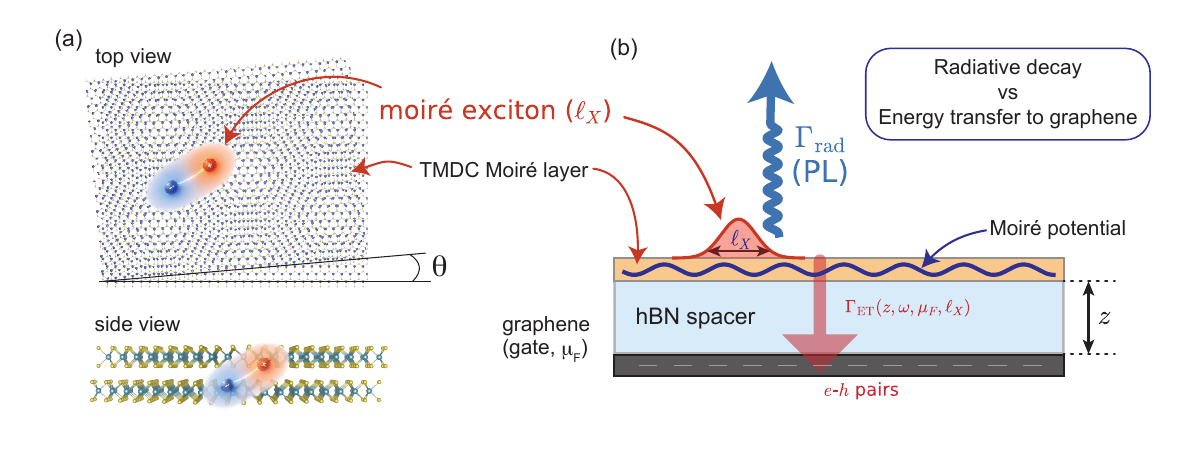}
  \caption{(Color online) (a) Moir\'e exciton---a bound electron--hole pair of center-of-mass
  extent $\lX$---in a transition-metal dichalcogenide (TMDC) heterobilayer twisted
  by angle $\theta$. Top view: the exciton trapped in the moir\'e superlattice;
  side view: the electron and hole occupy adjacent layers, forming an interlayer
  exciton. (b) Device cross section: the exciton in the TMDC moir\'e layer,
  confined by the moir\'e potential, can decay radiatively, emitting
  photoluminescence (PL) at rate $\Gamma_{\rm rad}$, or transfer its energy
  nonradiatively across a hexagonal boron nitride spacer of thickness $z$ to
  electron--hole pairs in a graphene gate at Fermi level $\muF$
  [$\GET(z,\omega,\muF,\lX)$], the channel studied here.}
  \label{fig:concept}
\end{figure*}

Two-dimensional (2D) semiconductors and their moir\'e superlattices provide a
qualitatively richer setting for this physics. The optical response of monolayer
and bilayer transition-metal dichalcogenides (TMDCs) is dominated by tightly bound
excitons with binding energies of hundreds of meV~\cite{Wang2018,Chernikov2014}.
In TMDC heterobilayers, interlayer excitons can be
trapped in the moir\'e potential, forming arrays of localized emitters whose
spatial extent $\lX$ ranges from a few to tens of
nanometers~\cite{Seyler2019,Tran2019,Jin2019,Shinokita2022,Kim2023,Ahmad2026};
the theory and broader
phenomenology of these moir\'e-trapped emitters are reviewed in
Refs.~\cite{Yu2017,Wu2018,MakShan2022}, and they act as gate-tunable quantum-light
sources~\cite{Baek2020}. These devices
commonly include nearby graphene or graphite layers acting as gates or
contacts, separated from the excitons by only a few nanometers of hexagonal
boron nitride (hBN) [Fig.~\ref{fig:concept}]. The same electrode used for
electrostatic control therefore sits within energy-transfer range of the excitons.

The nonradiative mechanism itself is general. It is not specific to moir\'e
excitons but operates for any emitter---a dye molecule, a quantum dot, a free or
disorder-localized monolayer exciton---placed near a 2D electronic
bath. What distinguishes these cases is the emitter's spatial extent, which
enters the transfer rate through a form factor $F_X(q)$ and, for a spatially
coherent emitter, through its total transition dipole $D_X$
(Table~\ref{tab:emitters}).
A point dipole has $F_X\to1$ and recovers the Swathi--Sebastian result. A
delocalized monolayer exciton has an effective $F_X$ set by coherence,
temperature, and disorder, none of them easily controlled. A moir\'e exciton is
the cleanest realization. Its center-of-mass wave function is localized by the
moir\'e potential~\cite{Wu2017}, so its localization length $\lX$ is an
\emph{intrinsic and tunable} parameter---set by twist angle, moir\'e-potential
depth, and registry.
Proximity quenching thus becomes a controllable probe of exciton
localization, and we take moir\'e excitons as the primary application of
an otherwise general formalism.

\begin{table*}[t]
  \caption{Emitters near a 2D electronic bath, classified by the (amplitude)
  exciton form factor $F_X(q)$. The rate~\eqref{eq:minimal} carries
  $|F_X(q)|^2$. The tabulated Gaussian form factors correspond to the coherent
  model, for which a spatially coherent exciton additionally carries a giant
  oscillator strength $|D_X|^2\propto\lX^2$. The fixed-dipole benchmark instead
  holds $D_X$ fixed and replaces $|F_X|^2=e^{-q^2\lX^2}$ by $e^{-q^2\lX^2/2}$.}
  \label{tab:emitters}
  \begin{ruledtabular}
  \begin{tabular}{lcl}
  emitter & $F_X(q)$ & localization \\
  \colrule
  point dipole / molecule & $1$ & none ($z^{-4}$ limit) \\
  finite-coherence exciton & effective & weak, uncontrolled \\
  disorder-localized & $e^{-q^2\ell_{\rm dis}^2/2}$ & random \\
  moir\'e exciton & $e^{-q^2\lX^2/2}$ & \textbf{tunable} \\
  \end{tabular}
  \end{ruledtabular}
\end{table*}

In this work we reformulate graphene-induced fluorescence quenching for the
modern context of moir\'e excitons. Related resonance-energy-transfer
problems---graphene near quantum dots, defect-bound single-photon emitters, or
moir\'e-trapped excitons---have been treated within the random-phase approximation
(RPA), which evaluates the transfer rate numerically for a specified emitter as a
function of its size, gate voltage, dielectric environment, and radiative
lifetime~\cite{Fezai2017,Eddhib2021,Hichri2021}. Our approach is instead analytic
and structural. We separate the total transition dipole $D_X$ from a normalized
transition form factor $F_X(q)$, and expand $F_X$ at the small momenta $q\sim1/z$
that the near-field filter admits. This expansion shows that, relative to the
point-dipole response of the same graphene bath, the \emph{leading} long-distance
correction depends on the emitter through only a single length---the
root-mean-square (rms) transition-polarization radius $R_X$.

We propose to extract $R_X$ from the normalized, spacer-thickness-dependent
quenching of a spectroscopically matched family of moir\'e-exciton lines, using
the graphene gate as an internal active/Pauli-blocked reference. The aim is thus to identify
\emph{which} geometric quantity the crossover measures, and under what conditions
the rich moir\'e structure reduces to it, rather than to compute a rate for one
specific system. We then delimit this minimal picture by benchmarking it against
the full RPA and by adding the anisotropic dielectric environment, the graphene
plasmon, and multilayer/graphite gates.

Our goals are threefold. First, we cast the
nonradiative ET rate $\GET(z,\omega,\muF,\lX)$ in a single transparent
expression---the overlap between the exciton near field and the electronic density
response of graphene---and verify that it contains the known $z^{-4}$ law as a
limit. Second, we show that the finite size of a moir\'e exciton, encoded in a
form factor $F_X(q)$, leaves a clear fingerprint of the transition-polarization
radius $R_X$---related to the localization length $\lX$ once a transition-source
profile is specified---in the distance dependence of the quenching. Third, we treat graphene as a gate-tunable bath, so that Pauli
blocking of its electron--hole continuum allows the nonradiative channel to be
electrically modulated. Throughout, we deliberately work in the long-wavelength
regime. Because a distance $z$ selects in-plane momenta $q\sim 1/z$, the essential
finite-size physics is captured by a controlled low-$q$ expansion, with the full
momentum-resolved RPA used as a benchmark.

\section{Theory}

\subsection{Golden-rule rate as a near-field/loss-function overlap}

The nonradiative ET rate follows from Fermi's golden rule,
\begin{equation}
  \GET=\frac{2\pi}{\hbar}\sum_f
  \bigl|\langle g;f_{\rm bath}|H_{\rm int}|e;0_{\rm bath}\rangle\bigr|^2
  \,\delta(E_f-E_0-\hbar\omega),
  \label{eq:golden}
\end{equation}
where $|e\rangle,|g\rangle$ are the excited and ground states of the exciton,
separated by the emission energy $\hbar\omega$, and
$|0_{\rm bath}\rangle,|f_{\rm bath}\rangle$ are the ground and excited states of
graphene. The emitter couples to the graphene electron density $\rho_G$ through
its transition Coulomb potential $\phi_X$,
\begin{equation}
  H_{\rm int}=\int d^2r\,\rho_G(\bm r)\,\phi_X(\bm r,z)
  =\sum_{\bm q}\rho_G(-\bm q)\,\phi_X(\bm q,z).
\end{equation}
The exciton is electrically neutral, so its transition charge density carries no
monopole; but its dipole and higher moments are finite, and it is this oscillating
transition dipole---not a net charge---that sources $\phi_X$ and drives the
transfer.
A near-field component of in-plane momentum $q$ decays away from the emitter as
$e^{-qz}$, so the transition probability carries a factor $e^{-2qz}$. Large-$q$
components are exponentially suppressed when the emitter is far from graphene,
and a given distance $z$ probes the electronic response around $q\sim 1/z$.

The sum over graphene final states in Eq.~\eqref{eq:golden} is the dynamic
structure factor $S(q,\omega)=\sum_f|\langle f|\rho_G(\bm q)|0\rangle|^2
\delta(E_f-E_0-\hbar\omega)$, which by the fluctuation--dissipation theorem is
proportional to the imaginary part of the graphene density response function
$\chi_G(q,\omega)$~\cite{Hwang2007}, with
$S(q,\omega)\propto[1+n_B(\omega)]\,[-\mathrm{Im}\,\chi_G(q,\omega)]$. At the
exciton energy $\hbar\omega\gg k_BT$ the Bose factor $1+n_B\simeq1$ and is
absorbed into the overall constant, distinct from the Pauli smearing of
Eq.~\eqref{eq:thermal} below. The rate therefore takes the
compact form
\begin{equation}
  \GET(z,\omega)\propto\int\frac{d^2q}{(2\pi)^2}\,
  |\phi_X(q,z)|^2\,\bigl[-\mathrm{Im}\,\chi_G(q,\omega)\bigr],
  \label{eq:overlap}
\end{equation}
i.e. the rate is the overlap between the emitter's near-field spectrum
$|\phi_X(q,z)|^2$ and the graphene density-response loss function
$L_G^{(0)}(q,\omega)\equiv-\mathrm{Im}\,\chi_G(q,\omega)$, its ability to absorb energy at
the same $(q,\omega)$ (the superscript $(0)$ marks undoped graphene; gate doping
enters below).
In this convention $\phi_X$ is the \emph{external} transition potential generated
by the exciton at the graphene plane, while $\chi_G$ is the density response of
graphene to that external potential. Screening internal to graphene is therefore
contained in $\chi_G$, not in $\phi_X$, which avoids double counting. The
corresponding screened-response kernel is made explicit in Sec.~\ref{sec:rpa}.

The golden rule~\eqref{eq:golden} treats the emitter as spectrally narrow, which is
accurate when $L_G^{(0)}$ varies slowly across the exciton linewidth. Near a sharp Pauli
edge or plasmon resonance, or for a narrow low-temperature emitter, the
$\delta$-function should be replaced by a convolution of $L_G^{(0)}$ with the exciton
lineshape. The same bath response also produces a dispersive exciton energy shift,
the real part of the self-energy
$\Sigma_X(\omega,z)\propto\int d^2q\,|\phi_X(q,z)|^2\chi_G(q,\omega)$ whose
imaginary part gives $\GET=-2\,\mathrm{Im}\,\Sigma_X/\hbar$. We keep this
dissipative part and neglect the Lamb-shift-like real part
$\delta E_X=\mathrm{Re}\,\Sigma_X$, which is largest near the Pauli edge, the
plasmon, or the shortest spacings and also complicates holding the exciton energy
fixed under gating.

\subsection{Exciton form factor and the minimal model}

For concreteness we first write the near-field factor for a perpendicular point
transition dipole, which produces the potential $\phi_X(q,z)\propto e^{-qz}$---the
dipole factor $q$ cancels the $1/q$ of the 2D Coulomb kernel---so
$|\phi_X|^2\propto e^{-2qz}$. An in-plane transition dipole gives the same
exponential $e^{-2qz}$ after angular averaging and differs only by an
orientation-dependent prefactor (Appendix~\ref{app:kernel}). The distance and
localization dependence discussed below are therefore insensitive to the precise
optical-dipole orientation. (Interlayer moir\'e excitons additionally carry a
\emph{static} out-of-plane electric dipole. It arises from the electron--hole
layer separation, which also reduces the optical overlap and lengthens the
radiative lifetime~\cite{Rivera2015}, and produces a linear Stark
shift~\cite{Ciarrocchi2019} that should be held fixed or compensated when the
graphene Fermi level is tuned. This static dipole is distinct from the optical
transition dipole considered here.)

A moir\'e exciton is instead spatially extended, and its coupling to graphene is
set by its transition-polarization density $\bm P_X(\bm R)$, a function of the
in-plane center-of-mass coordinate $\bm R$ measured from the local
moir\'e-potential minimum---the source of the near field (the
transition charge density is $\rho_X=-\nabla\cdot\bm P_X$).
Write $\bm P_X=\bm u\,P_X$, with $\bm u$ the unit vector along the transition
dipole. For an optically bright state [$P_X(\bm 0)\neq0$] we \emph{define} the total
transition dipole $D_X\equiv P_X(\bm 0)$ and the normalized form factor
$F_X(q)\equiv P_X(\bm q)/P_X(\bm 0)$, so that $P_X(\bm q)=D_X\,F_X(q)$ with
$F_X(0)=1$ identically. This split is exact and separates the oscillator strength
$D_X$ from a form factor that encodes the momentum filtering; the localization
length $\lX$ enters only through the single-envelope model below, where
$F_X(q;\lX)$ is set by the center-of-mass envelope and $D_X(\lX)$ acquires its
coherent-area scaling. The optical transition
polarization is moreover \emph{linear} in the center-of-mass
amplitude---within the envelope-function approximation,
$\bm P_X(\bm R)=\langle 0|\hat{\bm P}(\bm R)|X\rangle\propto\psi_X(\bm R)$~\cite{Matsuda2003}
(Appendix~\ref{app:formfactor}), with $\hat{\bm P}$ the polarization-density
operator---so the source profile
follows $\psi_X$, not the probability density $|\psi_X|^2$. For a normalized
Gaussian envelope $\psi_X(\bm R)\propto e^{-R^2/2\lX^2}$ this gives
(Appendix~\ref{app:formfactor})
\begin{equation}
  |F_X(q;\lX)|^2=e^{-q^2\lX^2},\qquad |D_X(\lX)|^2\propto\lX^2,
  \label{eq:formfactor}
\end{equation}
where the $\lX^2$ is the coherent (giant-oscillator-strength) enhancement that
arises from summing the transition dipole coherently over the localization
area~\cite{Rashba1962,NairTakagahara1997,Feldmann1987,Andreani1991,Hanamura1988}.

The microscopic optical factors---the relative-coordinate wave function at zero
electron--hole separation, interlayer overlap, and valley/spin selection
rules---set the local dipole prefactor, whereas the center-of-mass envelope fixes
both the coherent-area enhancement of $D_X$ and the momentum filtering through
$F_X$, suppressing components with $q\gtrsim 1/\lX$.

Two approximations underlie the identification of $F_X$ with the center-of-mass-envelope form factor.
First, we place the transition
polarization in a single effective 2D plane. For an interlayer exciton with
appreciable layer-resolved transition amplitudes the factor $e^{-qz}P_X(q)$
becomes a coherent sum $\sum_\alpha e^{-qz_\alpha}P_{X,\alpha}(q)$ over the
constituent layers $\alpha$, where $z_\alpha$ is the graphene-to-layer distance
and $P_{X,\alpha}$ the transition-polarization amplitude in layer $\alpha$. With
the layers separated by $\delta z\sim0.6$--$0.7$~nm, this vertical spread is
non-negligible ($e^{-q\delta z}\sim0.8$ at $q\sim0.4~\mathrm{nm}^{-1}$) once
$q\sim1/z$ grows.

Second, we treat the total transition dipole $D_X$ as $q$-independent, which holds
while the microscopic optical factors vary weakly over the near-field window
$q\sim1/z$. The relevant internal scale here is the electron--hole relative size
$a_X\sim1$--$2$~nm~\cite{Chernikov2014}---the internal exciton radius, distinct
from the center-of-mass envelope $\lX$. This approximation can fail at the smallest
spacings. When $q\sim1/z$ approaches $1/a_X$, the transition dipole acquires its
own microscopic form factor, $D_X\to D_X\,M_{\rm micro}(q)$ with
$M_{\rm micro}(0)=1$, set by the internal (relative-coordinate) exciton wave
function---the internal analogue of the envelope factor $F_X$. Both
$M_{\rm micro}$ and the vertical
(interlayer) factor above also suppress high $q$ and can therefore mimic an
additional finite lateral size in the graphene response.
Graphene therefore responds to the \emph{combined} transition-source form factor,
so isolating the lateral envelope radius $R_X$ requires the vertical and
microscopic factors to be known or negligible.

For this extended source the near-field potential is
$\phi_X(q,z)\propto D_X\,F_X(q;\lX)\,e^{-qz}$---the point-dipole factor $e^{-qz}$
times the source form factor $D_X F_X$---so that
$|\phi_X(q,z)|^2\propto|D_X|^2\,|F_X(q;\lX)|^2\,e^{-2qz}$. The neutral-graphene
interband response at small $q$ is $-\mathrm{Im}\,\chi_G\propto q^2/\omega$
(Appendix~\ref{app:kernel}). Inserting both into Eq.~\eqref{eq:overlap} with the
planar measure $d^2q=2\pi q\,dq$ reduces it to a single radial integral,
\begin{equation}
  \GET=C_G(\omega)\,|D_X(\lX)|^2\int_0^{\qc}\!dq\,
  q^3\,e^{-2qz}\,|F_X(q;\lX)|^2.
  \label{eq:minimal}
\end{equation}
Here $C_G(\omega)$ collects the graphene-response and dielectric-environment
prefactors---not the emitter oscillator strength, which is contained in
$|D_X|^2$. It is independent of $z$ and $\lX$ and therefore sets only the overall
scale.

The upper limit $\qc\simeq\omega/\vF$ is the kinematic cutoff above which a Dirac
electron--hole pair can no longer conserve energy and momentum. With the
graphene Dirac velocity $\hbar\vF=0.658~\mathrm{eV\,nm}$~\cite{CastroNeto2009} and
$\hbar\omega=1.5$~eV one has
$\qc\approx2.3~\mathrm{nm}^{-1}$ ($1/\qc\approx0.4$~nm). This sharp cutoff is only
a schematic stand-in for the kinematic edge. The exact response
[Eq.~\eqref{eq:imchi0}] has an integrable square-root singularity at $q=\qc$, but
the near-field factor $e^{-2qz}$ suppresses that edge exponentially, so for the
experimentally relevant spacings $z\gtrsim2$~nm---several times $1/\qc$, so that
$e^{-2\qc z}\lesssim10^{-4}$ at the cutoff---neither the cutoff nor the precise
edge shape affects the result.

A final filter is environmental rather than intrinsic to the exciton. Besides the
exciton's own size, the surrounding dielectric---the TMDC layers, the hBN spacer,
and their interfaces---also shapes the near field on its way to graphene. We
collect this into an \emph{electrostatic transfer function} $T(q,\omega)$, the
factor by which the stack transmits each momentum component of the potential to
graphene, so that $\phi_X\propto T(q,\omega)\,e^{-qz}P_X(q)$ and the rate
reads $\GET\propto\int q\,dq\,e^{-2qz}|T(q,\omega)|^2|F_X(q)|^2 L_G^{(0)}(q,\omega)$.
Microscopically, $T(q,\omega)$ carries the TMDC background
polarizability~\cite{Raja2017}, Keldysh-type 2D
screening~\cite{Keldysh1979,Cudazzo2011}, and interface and finite-slab
reflections. Because the measurement samples only $q\sim1/z$, what matters is how
$T(q,\omega)$ varies across that window. A flat $T(q,\omega)$ merely rescales the
rate. A $T(q,\omega)$ that falls with $q$ over $q\sim1/z$ instead suppresses the
high-$q$ near field
exactly as the exciton form factor $F_X(q)$ does, so this environmental
$q$-dependence can be misread as extra exciton size and bias the extracted $R_X$.
A Keldysh form $T(q,\omega)\sim(1+r_* q)^{-1}$, for instance, adds an $r_*/z$ term
of even lower order than the $R_X^2/z^2$ size correction. Any known
$T(q,\omega)$ must therefore be folded into the point-source kernel, as
$|T(q,\omega)|^2|F_X(q)|^2$, before $R_X$ is extracted. The minimal
kernel~\eqref{eq:minimal} takes $T(q,\omega)$ smooth over $q\sim1/z$ and absorbs
its $q\to0$ value into $C_G$.

For the extended coherent exciton the distance dependence is governed by
$|F_X(q;\lX)|^2=e^{-q^2\lX^2}$, while $|D_X|^2\propto\lX^2$ multiplies the whole
rate. Two consequences should be kept separate. For a \emph{fixed} exciton $D_X$
is a common multiplicative factor, so the normalized distance dependence and gate
modulation are independent of the oscillator-strength model, while the finite-size
deviation from the point-source curve is controlled by $F_X(q)$. Comparing
\emph{different} localization lengths additionally involves $D_X(\lX)$. As a
transparent benchmark that isolates momentum filtering at fixed oscillator
strength, we also use a fixed-total-dipole model in which $D_X$ is held constant
and the source is a normalized Gaussian transition-polarization profile, so that
$|F_X(q;\lX)|^2=e^{-q^2\lX^2/2}$. For this Gaussian parametrization the profile
width coincides numerically with that of $|\psi_X|^2$, but it is not interpreted as
a probability-density source. This benchmark recovers the point-dipole $z^{-4}$ law
for every $\lX$ at large $z$.

At fixed total transition dipole $D_X$, the form-factor kernel approaches the
point-dipole result as $\lX\to0$. The point-dipole molecule of Swathi and
Sebastian ($F_X\to1$) recovers, for $\qc\to\infty$,
\begin{equation}
  \GET\propto\int_0^\infty dq\,q^3 e^{-2qz}=\frac{3}{8z^4},
  \label{eq:z4}
\end{equation}
the $z^{-4}$ law of Refs.~\cite{Swathi2008,Swathi2009}. In the coherent-envelope
model, by contrast, $|D_X|^2\propto\lX^2$, so the mathematical limit $\lX\to0$
sends $\GET\to0$ and does not describe a point molecule. It lies outside the
weak-confinement envelope description. For $\lX>0$ the Gaussian factor removes the
high-$q$ part of the near field, so that the rate falls below its own $z^{-4}$
asymptote once $z\lesssim\lX$.

\subsection{General form factor: low-$q$ expansion and the moir\'e reduction}
\label{sec:lowq}

The Gaussian envelope above is a convenient special case. The near-field filter
makes the finite-size physics far more general. For an arbitrary
scalar transition-polarization amplitude $p_X(\bm R)$---the source magnitude
introduced above---with $D_X=\int d^2R\,p_X$ and
$F_X(q)=D_X^{-1}\int d^2R\,p_X(\bm R)\,e^{-i\bm q\cdot\bm R}$
(Appendix~\ref{app:formfactor}), centering the source ($\int d^2R\,\bm R\,p_X=0$)
and Taylor-expanding the plane-wave factor $e^{-i\bm q\cdot\bm R}$ in small $q$
gives, for an isotropic bright state,
\begin{equation}
\begin{gathered}
  F_X(q)=1-\tfrac14 q^2\langle R^2\rangle_P+O(q^4),\\
  \langle R^2\rangle_P\equiv\frac{\int d^2R\,R^2 p_X(\bm R)}{\int d^2R\,p_X(\bm R)},
\end{gathered}
  \label{eq:lowq}
\end{equation}
where $\langle R^2\rangle_P$ is the second radial moment of the source. For a
nodeless, phase-uniform $s$-like bright state we fix the global phase so that
$D_X>0$ and $p_X(\bm R)$ is real. The second moment $\langle R^2\rangle_P$ then
defines a positive transition-polarization radius $R_X\equiv\langle R^2\rangle_P^{1/2}$.
This is the second-moment radius of the optical transition polarization. It need
not equal the rms radius of the center-of-mass probability density $|\psi_X|^2$,
and converting between the two requires a source model.

It follows that $|F_X(q)|^2=1-\tfrac12 q^2\langle R^2\rangle_P+O(q^4)$, and inserting
this into the rate gives the leading long-distance correction to the point-dipole law,
\begin{equation}
  \GET(z)=\frac{3C_G|D_X|^2}{8z^4}
  \left[1-\frac{5}{2}\frac{\langle R^2\rangle_P}{z^2}
  +O\!\left(\frac{\langle R^4\rangle_P}{z^4}\right)\right],
  \label{eq:leadingR}
\end{equation}
valid for $z\gg R_X,\qc^{-1}$, where the next-order term involves the
\emph{independent} fourth moment $\langle R^4\rangle_P$.
Within the minimal $q^2$ bath kernel, this \emph{first} emitter-induced deviation
from the $z^{-4}$ law is governed by $R_X^2$ alone, independently of the detailed
source, with higher moments entering only in the full crossover at $z\sim R_X$.
More generally, this universality applies to the ratio to the point-dipole
response evaluated with the same bath. For the Gaussian envelope
$\langle R^2\rangle_P=2\lX^2$ (coherent) or $\lX^2$ (fixed-dipole benchmark),
consistent with Eq.~\eqref{eq:formfactor}, so in $R_X$ the two Gaussian form
factors coincide and their crossover curves collapse, with a half-suppression at
$z\simeq1.6\,R_X$ (Sec.~\ref{sec:size}). For a general non-Gaussian source this
collapse and half-suppression are not universal---only the leading deviation is.
Equation~\eqref{eq:leadingR} therefore gives a model-independent $R_X$ only for
$z\gg R_X$; the stronger crossover requires the transition-polarization profile or
higher moments.

This two-term expansion converges slowly. For a Gaussian source the exact
ratio $\GET/\GET^{\rm pd}$ of the finite-size rate to the point-dipole rate
$\GET^{\rm pd}$ ($F_X\to1$ at the same $D_X$) is $0.62$, $0.82$, and $0.91$ at
$z/R_X=2$, $3.3$, and $5$, whereas Eq.~\eqref{eq:leadingR} gives $0.38$, $0.78$,
and $0.90$. The leading
correction is thus quantitative only for $z/R_X\gtrsim4$--$5$. In the crossover
$z\sim R_X$, where the signal is largest, the full form factor or a controlled
multi-moment expansion must be used. The next term is
$[\tfrac{105}{32}\langle R^2\rangle_P^2+\tfrac{105}{64}\langle R^4\rangle_P]/z^4$,
reducing for a Gaussian to $+\tfrac{105}{16}R_X^4/z^4$.

This universality also presumes a fixed bath response. The bare graphene kernel is
itself weakly $q$-dependent. Expanding Eq.~\eqref{eq:imchi0} gives
$-\mathrm{Im}\,\chi_0\propto(q^2/\omega)[1+\tfrac12(\vF q/\omega)^2+\cdots]$, which
adds a bath term $+\tfrac52\lambda_\omega^2/z^2$ inside the bracket of
Eq.~\eqref{eq:leadingR}, with $\lambda_\omega\equiv\hbar\vF/\hbar\omega\simeq0.44$~nm
at $\hbar\omega=1.5$~eV. The raw deviation from a pure $z^{-4}$ power law therefore
carries $\tfrac52(\lambda_\omega^2-R_X^2)/z^2$, whereas the ratio to the
point-dipole rate evaluated with the same bath,
$\GET(z)/\GET^{\rm pd}(z)=1-\tfrac52 R_X^2/z^2+\cdots$ (with $z\to z_{\rm eff}=\gamma z$
once the spacer-anisotropy factor $\gamma$ of Sec.~\ref{sec:aniso} is included),
cancels the bath term and isolates $R_X$. The bath correction is $\sim2\%$ of the size term for $R_X=3$~nm
but grows to $\sim10$--$20\%$ for $R_X\lesssim1$--$1.5$~nm, so $R_X$ is best
extracted from a fit that retains the known graphene kernel---equivalently the
ratio to the point-dipole rate---rather than from the deviation from a bare
$z^{-4}$ law.

The single-radius reduction presumes an isolated optically bright state
[$D_X=P_X(0)\neq0$] and does not apply to dark, nodal, or phase-textured states,
whose coupling is instead set by the leading nonvanishing multipole. Because
graphene accepts finite in-plane momentum, it can moreover couple to
momentum-dark excitons [$P_X(0)=0$, $P_X(\bm Q)\neq0$] that are invisible in
far-field PL. Their graphene-induced depletion can affect a bright line even at
fixed $R_X$, but requires an unnormalized finite-$q$, multilevel treatment
deferred here.

Moir\'e excitons carry microscopic structure that this expansion makes explicit.
Physically, the transition polarization is the slow center-of-mass envelope
$\psi_X(\bm R)$ times a \emph{local} transition dipole $\bm d(\bm R)$---a vector
field whose strength, direction, and valley/spin character follow the atomic
registry and therefore repeat with the moir\'e period. Its Fourier series over the
moir\'e reciprocal lattice, $\bm d(\bm R)=\sum_{\bm G_M}\bm d_{\bm G_M}\,e^{i\bm G_M\cdot\bm R}$,
gives
\begin{equation}
\begin{gathered}
  \bm P_X(\bm R)=\psi_X(\bm R)\sum_{\bm G_M}\bm d_{\bm G_M}\,e^{i\bm G_M\cdot\bm R},\\
  \bm P_X(\bm q)=\sum_{\bm G_M}\bm d_{\bm G_M}\,\tilde\psi_X(\bm q-\bm G_M),
\end{gathered}
  \label{eq:moire}
\end{equation}
an envelope-times-Bloch (wave-packet) form in which the coefficients
$\bm d_{\bm G_M}$ encode the registry-dependent optical selection rules, valley and
spin structure, and interlayer hybridization at each moir\'e reciprocal vector
$\bm G_M$~\cite{Wu2018,Yu2017}, while $\tilde\psi_X$ is the center-of-mass envelope.
The cell average $\bm d_{\bm 0}$ of $\bm d(\bm R)$---the coarse-grained, bright
dipole---supplies the microscopic bright-dipole factor; in this reduction the
total transition dipole is $D_X=|\bm d_{\bm 0}|\,\tilde\psi_X(\bm 0)$, including the
coherent-envelope contribution $\tilde\psi_X(\bm 0)\propto\lX$ that yields the
giant oscillator strength $|D_X|^2\propto\lX^2$. The higher harmonics
$\bm d_{\bm G_M\neq\bm0}$ describe how $\bm d(\bm R)$ varies within a moir\'e cell.

For a bright, $s$-like ground state, two conditions make the $\bm G_M\neq\bm0$
(Umklapp) terms negligible at the sampled momenta $q\sim1/z$. First,
$z|\bm G_M|\gg1$: the near-field filter $e^{-qz}$ passes only $q\sim1/z\ll|\bm G_M|$,
so components displaced by a nonzero $\bm G_M$ are exponentially suppressed.
Second, $\lX|\bm G_M|\gg1$: an envelope spread over many moir\'e cells
($\lX\gg1/|\bm G_M|$) is correspondingly narrow in momentum (width
$\sim1/\lX\ll|\bm G_M|$), so the copies $\tilde\psi_X(\bm q-\bm G_M)$ centered at
each $\bm G_M$ do not overlap, and those with $\bm G_M\neq\bm0$ leave negligible
weight near $q\sim0$. When both hold, only the $\bm G_M=\bm 0$ term
survives and
$\bm P_X(\bm q)\simeq\bm d_{\bm G=\bm 0}\,\tilde\psi_X(\bm q)\equiv D_X F_X(q)$.
The microscopic moir\'e physics enters through the prefactor $D_X$, whereas the
low-$q$ distance dependence is governed by the envelope radius $R_X$ alone.

The reduction further requires the bright $\bm G_M=\bm 0$ coefficient to dominate. It
holds only when $|\bm d_{\bm G_M}\tilde\psi_X(\bm q-\bm G_M)|\ll|\bm d_{\bm 0}\tilde\psi_X(\bm q)|$
across the near-field window $q\sim1/z$, which can fail for registries with a small
$\bm d_{\bm 0}$ even when the envelope factor is exponentially suppressed. This
is the precise sense in which a moir\'e exciton reduces to a single length scale,
but the conditions are not automatic. They hold for typical short-period moir\'es
with sufficiently smooth ground-state envelopes. A moir\'e period $a_M=10$~nm gives
$|\bm G_M|=4\pi/(\sqrt3\,a_M)\simeq0.73~\mathrm{nm}^{-1}$, so the Umklapp weight
$e^{-2|\bm G_M|z}$ is $\sim7\times10^{-4}$ at $z=5$~nm and $\sim10^{-2}$ at
$z=3$~nm, while $\lX|\bm G_M|\simeq7$ for $\lX=10$~nm but only $\simeq2$ for
$\lX=3$~nm. The reduction can fail, however, for long-period moir\'es or very thin
spacers ($z\lesssim2$~nm), and for tightly localized, excited, phase-textured,
momentum-dark, or strongly registry-modulated states---cases in which the
$\bm G_M\neq0$ terms, higher envelope moments, or the full transition density must
be retained rather than a single $R_X$.

The scalar reduction further assumes a single bright polarization channel with a
common dipole direction $\bm u$. Multiple or spatially varying channels replace
$F_X$ and $\langle R^2\rangle_P$ by the corresponding tensors.
Equation~\eqref{eq:moire} is itself a single-envelope ansatz. A microscopic
continuum model can entangle the moir\'e harmonics, layer composition, and
registry-dependent optical dipole into a superposition over several harmonics
whose transition density does not factor into a smooth envelope times a
$\bm G_M=0$ dipole, in which case the full $P_X(q)$ must be retained. Such a
state-resolved microscopic evaluation is beyond the universal long-wavelength
theory developed here.

In summary, the reduction of the graphene coupling to a single
transition-polarization radius $R_X$ rests on several conditions: an isolated,
phase-uniform bright state; a transition polarization effectively confined to one
plane; a microscopic and environmental transfer function that varies slowly over
$q\sim1/z$; negligible gate- or spacer-induced modification of the exciton state;
and, for the extraction protocol below, approximately single-exponential population
decay. When these hold, proximity quenching measures the low-$q$
transition-polarization form factor, and its leading finite-size correction is
governed by $R_X$ alone. When they fail, the full layer-, momentum-, and
state-resolved transition density must be retained, and the single-$R_X$ reduction
becomes a hypothesis to be tested against a microscopic model.

\subsection{Gate tunability: Pauli blocking}

Doping graphene to a Fermi level $\muF$ blocks interband transitions whose final
state is already occupied. A vertical interband Dirac transition of energy
$\hbar\omega$ requires an empty final state, available only when
$\hbar\omega>2|\muF|$. This gate-tunable Pauli blocking of the interband
absorption is well documented in graphene optical
spectroscopy~\cite{Wang2008,Li2008}. As a first step we adopt the sharp
Pauli-blocking model
\begin{equation}
  L_G(q,\omega,\muF)=L_G^{(0)}(q,\omega)\,
  \Theta\!\left(\hbar\omega-2|\muF|\right),
  \label{eq:pauli}
\end{equation}
where $L_G^{(0)}(q,\omega)$ is the neutral-graphene loss function of
Eq.~\eqref{eq:overlap} and $L_G(q,\omega,\muF)$ its Pauli-blocked form (the extra
argument $\muF$ marks the doped response).

The exact interband threshold tilts with momentum, $\hbar\omega>2|\muF|-\hbar
\vF q$. The near-field filter selects $q\sim1/z$, which is smaller than, but not
always negligible compared with, the Fermi momentum $k_F$. In the optical-limit
approximation the blocking factor is taken to be $q$-independent and multiplies
$\GET$ as a whole, suppressing the interband contribution above $|\muF|=\hbar\omega/2$. The momentum tilt is not
entirely negligible at the shortest spacings, however. Blocking the representative
near-field momentum $q_{\rm peak}=3/2z$ (where the rate weight $q^3e^{-2qz}$ peaks)
requires $|\muF|\gtrsim(\hbar\omega+\hbar\vF q_{\rm peak})/2$, i.e. $\approx0.85$ and
$0.91$~eV at $z=5$ and $3$~nm (versus $0.75$~eV at $q=0$), so the sharp $q=0$
threshold underestimates the gate voltage needed for complete blocking at short
range, as the full momentum-resolved RPA polarizability of
Sec.~\ref{sec:rpa} confirms.

At finite temperature the interband
edge is thermally smeared. The $q\!\to\!0$ (optical) graphene interband response
gives the rigorous form~\cite{Falkovsky2007}
\begin{equation}
  \Theta(\hbar\omega-2|\muF|)\;\longrightarrow\;
  \frac12\sum_{\pm}\tanh\frac{\hbar\omega\pm2|\muF|}{4k_BT},
  \label{eq:thermal}
\end{equation}
which reduces to the step as $T\to0$ and broadens the threshold over
$\sim 4k_BT$ ($\approx0.1$~eV at room temperature). A more complete treatment
would use the full Dirac-electron RPA
response~\cite{Wunsch2006,Hwang2007}, including the intraband (Drude/plasmon)
channel. We develop this in Sec.~\ref{sec:rpa}.

\subsection{From rates to PL observables}

The total exciton decay rate is
$\Gamma_{\rm tot}=\Gamma_{\rm rad}+\Gamma_{\rm nr}^0+\GET$, with $\Gamma_{\rm
rad}$ the radiative rate and $\Gamma_{\rm nr}^0$ the intrinsic (graphene-free)
nonradiative rate. The excited-state population decays at the total rate
$\Gamma_{\rm tot}$---every channel, radiative or not, removes an
exciton---so the PL lifetime is $\tau_{\rm PL}=1/\Gamma_{\rm tot}$, whereas
only a fraction $\Gamma_{\rm rad}/\Gamma_{\rm tot}$ of those decays emit a
photon; the PL intensity is therefore proportional to this quantum yield. Here $I_{\rm PL}$ is the spectrally and
temporally integrated PL intensity under fixed excitation and collection
conditions. In a pulsed (time-resolved) measurement the PL decays as
$I_{\rm PL}(t)\propto\Gamma_{\rm rad}N_0\,e^{-\Gamma_{\rm tot}t}$, with $N_0$ the
pulse-created initial population: the graphene channel raises
$\Gamma_{\rm tot}$ and thus shortens the lifetime $\tau_{\rm PL}$, but leaves
the $t=0$ peak height $\Gamma_{\rm rad}N_0$ unchanged at fixed
$N_0$. In a decay trace, graphene quenching then appears as a faster decay and a
smaller area under the curve, but not as a lower initial height at $t=0$.

The absolute transfer rate carries an overall microscopic prefactor that we
prefer not to fix. We therefore measure spacings in units of the \emph{graphene
quenching distance} $d_0$, the graphene analogue of the F\"orster radius: the
spacing at which the \emph{point-dipole} transfer rate $\GET^{\rm pd}$---evaluated
with the same total transition dipole $D_X$ ($F_X\to1$)---equals the emitter's
intrinsic (graphene-free) decay rate,
$\GET^{\rm pd}(d_0)=\Gamma_0(\lX)\equiv\Gamma_{\rm rad}(\lX)+\Gamma_{\rm nr}^0(\lX)$.
At this spacing a point emitter's total rate is doubled,
$\Gamma_{\rm tot}=\Gamma_0+\GET^{\rm pd}=2\Gamma_0$, so its lifetime
($\tau_{\rm PL}=1/\Gamma_{\rm tot}$) and integrated PL ($\propto1/\Gamma_{\rm
tot}$) fall to half their graphene-free values. Referencing distances to $d_0$ absorbs the overall constant
entirely, so finite-size effects enter only as corrections relative to it.

The value $d_0\sim15$~nm used in the figures is illustrative: it sets the overall
scale and is of the same order as the graphene energy-transfer distances measured
for quantum emitters~\cite{Gaudreau2013,Tielrooij2015}. All
trends below rescale for a different,
experimentally determined $d_0$, so the predictions are statements about scaling
and device design rather than about an absolute rate. Writing
$x\equiv\GET/\Gamma_0$, both observables, measured relative to the graphene-free
emitter, reduce to the same function
\begin{equation}
  \frac{\tau_{\rm PL}}{\tau_0}=\frac{I_{\rm PL}}{I_0}=\frac{1}{1+x},
  \qquad
  x=\frac{\GET(z,\lX,\muF)}{\Gamma_0(\lX)}.
  \label{eq:pl}
\end{equation}
Here $\tau_0=1/\Gamma_0$ and $I_0$ are the lifetime and integrated PL intensity
of the same emitter without graphene.

The relative PL intensity and lifetime should therefore coincide whenever only
$\GET$ varies---a directly testable prediction. Equation~\eqref{eq:pl} assumes the radiative rate
$\Gamma_{\rm rad}$ is unchanged by graphene. Graphene does modify
$\Gamma_{\rm rad}$ by reshaping the emitter's photonic local density of states;
capturing that shift would require the full electromagnetic dyadic Green's
function of the layered stack---the object that fixes a dipole's decay rate near
any structure. In the strong-quenching regime considered here we assume that this
near-field nonradiative channel $\GET$ dominates over the graphene-induced
radiative-LDOS correction, which we therefore omit. Equation~\eqref{eq:pl} is
applied with a common $d_0$ for variations of $z$ or $\muF$ at \emph{fixed}
$\lX$, for which $D_X$, $\Gamma_{\rm rad}$, and $\Gamma_{\rm nr}^0$ are unchanged.
When different localization lengths are compared, the reference $\Gamma_0(\lX)$
and $d_0(\lX)$---and hence the oscillator strength---must be calibrated separately
for each exciton state.

\section{Results}
\label{sec:results}

All curves below use $\hbar\omega=1.5$~eV (a representative TMDC interlayer-exciton
energy), $\hbar\vF=0.658~\mathrm{eV\,nm}$, and, where a scale is needed, a
quenching distance $d_0=15$~nm and room temperature $T=300$~K for the interband
edge [Eq.~\eqref{eq:thermal}].

\subsection{Benchmark: recovery of the $z^{-4}$ law}

\begin{figure}[t]
  \centering
  \includegraphics[width=\columnwidth]{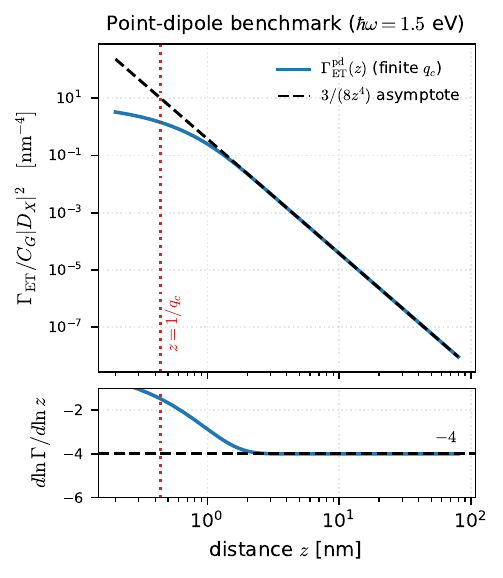}
  \caption{(Color online) Benchmark for a point dipole ($F_X=1$, fixed $D_X$). Top: ET rate (solid)
  and the $3/(8z^4)$ asymptote (dashed); the vertical line marks $z=1/\qc$.
  Bottom: local logarithmic slope, approaching $-4$ at large $z$.}
  \label{fig:bench}
\end{figure}

Figure~\ref{fig:bench} shows the point-dipole rate
$\GET^{\rm pd}(z)=C_G(\omega)\,|D_X|^2\int_0^{\qc}dq\,q^3e^{-2qz}$. On a log--log scale it follows a
slope of $-4$ over the nanometric, nonretarded range relevant here. A power-law fit for
$z>20$~nm gives an exponent of $-4.0000$, confirming that the numerical
evaluation recovers the analytic limit Eq.~\eqref{eq:z4}
of Refs.~\cite{Swathi2008,Swathi2009,Gaudreau2013}. The finite cutoff
$\qc$ bends the curve below the $z^{-4}$ asymptote only for $z\lesssim 1/\qc\sim
0.4$~nm, far below any realistic hBN spacer, so the cutoff plays no role in
practice.

\subsection{Finite-size effect: distance dependence as a probe of localization}
\label{sec:size}

\begin{figure*}[t]
  \centering
  \includegraphics[width=\textwidth]{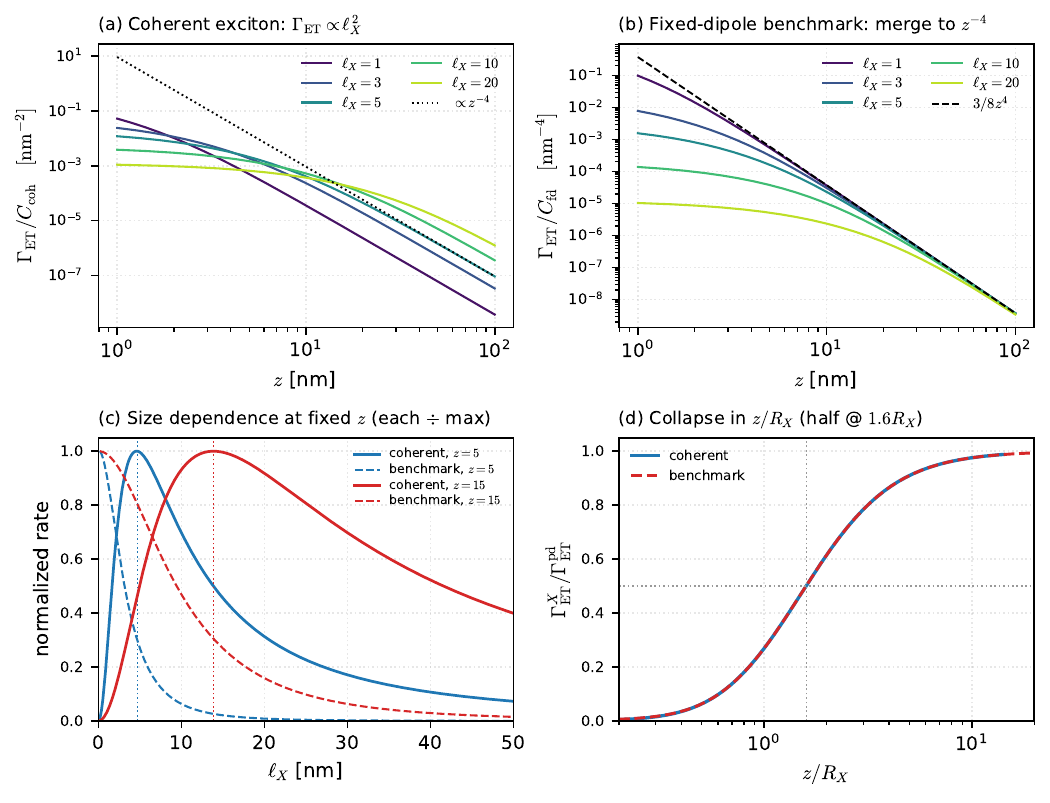}
  \caption{(Color online) Finite-size effect: coherent exciton versus fixed-dipole benchmark.
  (a) Coherent model ($|D_X|^2\propto\lX^2$, $|F_X|^2=e^{-q^2\lX^2}$): the rate
  (in units of the constant coherent prefactor $C_{\rm coh}$) splits by $\lX^2$
  and the curves cross---small excitons dominate at short range, large excitons at
  long range---while every curve keeps the $z^{-4}$ slope (dotted). (b)
  Fixed-dipole benchmark ($D_X$ const, prefactor $C_{\rm fd}$): all curves merge
  onto the point-dipole $3/(8z^4)$ asymptote (dashed). (c) Size dependence at fixed $z$
  (each curve normalized to its own maximum): the coherent rate is
  \emph{non-monotonic}, peaking near $\lX\approx z$ (dotted verticals), whereas
  the benchmark decreases monotonically. (d) Momentum-filtering shape (rate
  divided by the point-dipole rate evaluated with the same bath,
  $\GET^{X}/\GET^{\rm pd}$) versus $z/R_X$, with
  $R_X=\langle R^2\rangle_P^{1/2}$ the rms transition-polarization radius
  ($R_X=\lX$ and $\sqrt2\,\lX$ for the two models): expressed in $R_X$ the two
  Gaussian curves collapse, with a half-suppression at $z\simeq1.6\,R_X$; this
  exact collapse is specific to the two Gaussian models, whereas for a general
  source only the leading long-distance coefficient is universal.}
  \label{fig:size}
\end{figure*}

Figure~\ref{fig:size} contrasts the two source models. The coherent exciton is
the physical case; the fixed-dipole benchmark is a deliberately different foil,
included to separate the model-independent content (the extraction of $R_X$) from
the fingerprint of the coherent giant oscillator strength (the absolute
$\lX$-dependence of the rate). In both, $\GET(z)$ tracks
the point-dipole $z^{-4}$ law at large spacing ($z\gg\lX$), where the emitter
looks point-like. Once the spacing drops below the exciton size ($z\lesssim\lX$),
the finite source suppresses the short-wavelength ($q\sim1/z$) near field and the
rate flattens, deviating from $z^{-4}$. This distance-sweep
crossover is the robust signature of the transition-polarization radius. Because it
comes from the shape $|F_X(q)|^2$ and not from $D_X$, it is independent of the
oscillator-strength model.

The
half-suppression point---where the rate falls to half of the point-dipole rate
evaluated with the same bath---sits at $z\approx1.6\,\lX$ for the fixed-dipole benchmark and
$z\approx2.25\,\lX$ for the coherent exciton. This apparent
model dependence is an artifact of the nominal $\lX$. The two Gaussian models
assign different transition-polarization radii to the same $\lX$
[$\langle R^2\rangle_P=\lX^2$ and $2\lX^2$, Eq.~\eqref{eq:lowq}], and because in
the physical radius $R_X=\langle R^2\rangle_P^{1/2}$ their form factors coincide,
both collapse onto the same curve with a half-suppression at $z\approx1.6\,R_X$
[panel (d)]. What is universal for
an \emph{arbitrary} source, however, is only the leading long-distance deviation
$\propto R_X^2$ [Eq.~\eqref{eq:leadingR}], while the full crossover can involve
higher moments. The clean, model-independent observable is therefore $R_X$, extracted from the
leading large-$z$ deviation from the point-dipole response evaluated with the same
bath, rather than the convention-dependent envelope length $\lX$.

In the \emph{absolute} rate, by contrast, the two models differ in their
dependence on $\lX$. In
the fixed-dipole benchmark [panel (b)] the curves merge onto a common $z^{-4}$
asymptote and $\GET$ decreases monotonically with $\lX$. In the coherent model
[panel (a)] the giant oscillator strength $|D_X|^2\propto\lX^2$ makes the
long-range amplitude grow with $\lX$, so the curves separate by $\lX^2$ and cross.
Small excitons quench more strongly at short range, large excitons at long range.
At fixed distance this produces a \emph{non-monotonic} size dependence
[panel (c)], with $\GET(\lX)$ peaking near $\lX\approx z$, a distinctive
prediction of the coherent picture. This model dependence of the absolute rate shapes how one should measure the
exciton size. Comparing excitons of \emph{different} $\lX$ requires the
oscillator-strength law $D_X(\lX)$, which differs between the fixed-dipole and
fully-coherent limits and so cannot cleanly isolate the size. Sweeping the spacing
$z$ for a \emph{fixed} transition source, by contrast, leaves $D_X(\lX)$ out
entirely: the crossover position measures $R_X$ directly. And because a moir\'e exciton's
$\lX$ is intrinsic and twist-tunable---unlike the fixed size of a quantum
dot~\cite{Fezai2017}---this distance sweep becomes a tunable probe of $R_X$ (and,
once a source model is chosen, of $\lX$).

\subsection{Gate dependence: Pauli-blocked bath}

\begin{figure*}[t]
  \centering
  \includegraphics[width=\textwidth]{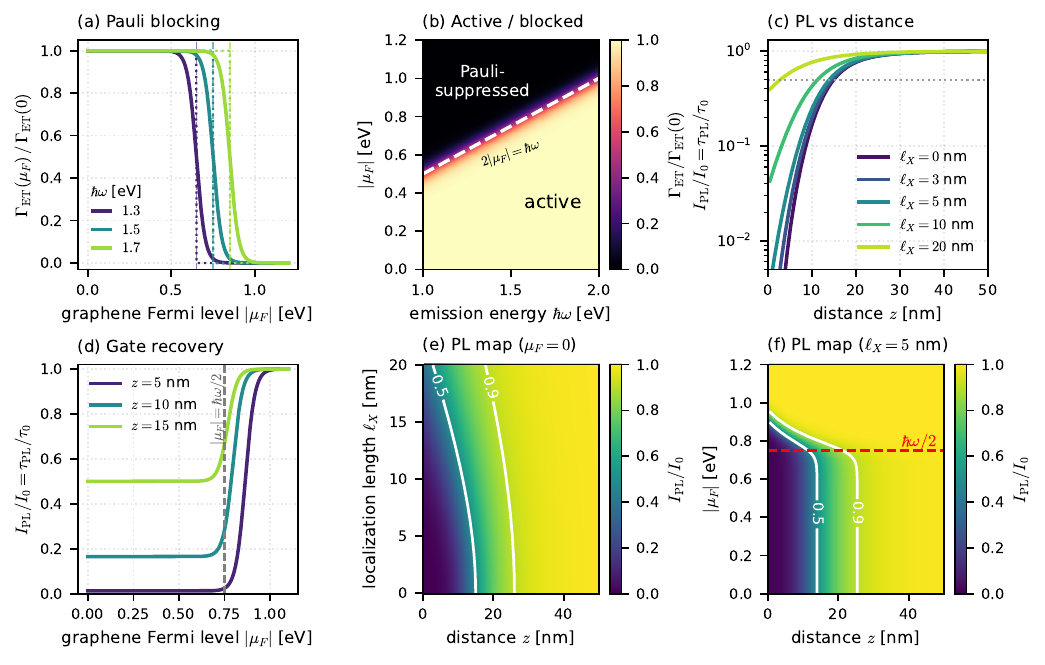}
  \caption{(Color online) Gate tunability and PL observables. (a) ET
  suppression factor versus $|\muF|$ for three emission energies (solid:
  $T=300$~K; dotted: $T=0$ step; vertical dashed: $|\muF|=\hbar\omega/2$); (b)
  active versus Pauli-blocked regions in the $(\hbar\omega,|\muF|)$ plane; (c)
  relative PL $I_{\rm PL}/I_0=\tau_{\rm PL}/\tau_0$ versus distance (log scale) for several
  $\lX$ at $\muF=0$ (fixed-dipole benchmark, for which $R_X=\lX$); (d) relative PL
  versus Fermi level for a point dipole ($\lX=0$) at several fixed distances,
  showing gate-induced recovery; (e) PL map in
  $(z,\lX)$ at $\muF=0$ (fixed-dipole benchmark); (f) PL map in $(z,|\muF|)$ at
  $\lX=5$~nm (red dashed:
  $|\muF|=\hbar\omega/2$; white contours $0.5,0.9$). Panels (b), (d), and (f) use
  the sharp $q=0$ optical-limit blocking factor; the quantitative finite-$q$ gate
  dependence is given by the full RPA in Fig.~\ref{fig:rpa}.}
  \label{fig:gatepl}
\end{figure*}

Figure~\ref{fig:gatepl}(a) shows the suppression of $\GET$ with graphene Fermi
level. In the minimal model the interband channel is strongly suppressed as
$2|\muF|$ approaches $\hbar\omega$, i.e. $|\muF|\to\hbar\omega/2$ (0.65, 0.75, and
0.85~eV for $\hbar\omega=1.3,1.5,1.7$~eV), thermally broadened over $\sim4k_BT$ at
finite temperature [Eq.~\eqref{eq:thermal}]. Residual intraband absorption,
disorder, and many-body effects make the suppression strong but not complete in
practice. The $(\hbar\omega,|\muF|)$ map [panel (b)] separates an active region,
where graphene is a strong quencher, from a Pauli-blocked region where its
interband absorption is strongly suppressed. This is the mechanism behind
electrically controlled relaxation
pathways~\cite{Tielrooij2015}, here applied to moir\'e excitons.

\subsection{PL observables and parameter regimes}

Translating to observables through Eq.~\eqref{eq:pl}, Fig.~\ref{fig:gatepl}(c)
shows the relative PL versus distance: at $z=5$~nm and $\muF=0$ a point-dipole
emitter is strongly quenched ($I_{\rm PL}/I_0\approx0.01$), while a finite exciton
recovers---for the fixed-dipole benchmark with $\lX=10$~nm, to $\approx0.16$---because
its form factor suppresses the short-wavelength (high-$q$) near field that couples
it to graphene. Figure~\ref{fig:gatepl}(d) shows
the same ratio versus gate voltage: gating graphene past $|\muF|=\hbar\omega/2$
partially restores the PL even at fixed distance, an independent electrical means
of distinguishing graphene-induced quenching from disorder.

Figure~\ref{fig:gatepl}(e,f) maps the relative PL $I_{\rm PL}/I_0$ over the device
parameter space. In the $(z,\lX)$ plane [panel (e)], the contour $I_{\rm PL}/I_0=0.9$ at
which quenching becomes negligible lies near $z\approx24$--$26$~nm and depends
only weakly on $\lX$, while strong quenching ($<0.5$) dominates below
$z\sim10$--$15$~nm. In the $(z,|\muF|)$ plane [panel (f)], gating above
$\hbar\omega/2$ suppresses the quenching, most completely at the larger spacings;
the quantitative, momentum-resolved gate requirement follows from the full RPA
[Fig.~\ref{fig:rpa}(c,d)]. These maps delineate when
proximity-induced ET must be included in the interpretation of moir\'e-exciton
PL.

One caveat on reading the $\lX$-varying panels [(c),(e)]: they are drawn with the
benchmark, whose $\lX$-independent $\Gamma_0$ lets one $d_0$ normalize the whole
family, whereas the fixed-$\lX$ panels [(b),(d),(f)] carry $D_X$ only as a constant
and are model-independent. For the physical coherent exciton the giant oscillator
strength puts $|D_X|^2\propto\lX^2$ [Eq.~\eqref{eq:formfactor}] into \emph{both}
$\Gamma_{\rm rad}$ and $\GET$. For a radiatively limited emitter
($\Gamma_0\simeq\Gamma_{\rm rad}\propto\lX^2$) this common factor cancels in
$x=\GET/\Gamma_0$; the benchmark curves then apply once the size is expressed
through the transition-polarization radius $R_X$, since $R_X=\sqrt2\,\lX$ for the
coherent Gaussian but $R_X=\lX$ for the benchmark. If instead $\Gamma_{\rm nr}^0$
dominates ($\Gamma_0$ nearly $\lX$-independent), an additional $\lX^2$ dependence
remains in the absolute quenching level, while the normalized momentum-filtering
crossover is still governed by $R_X$.

\emph{Worked example.} To make these trends concrete, consider a
TMDC heterobilayer with a moir\'e exciton of transition-polarization
radius $R_X=3$~nm separated from a graphene gate by an hBN spacer, taking
$\hbar\omega=1.5$~eV and the representative $d_0=15$~nm. At $z=5$~nm and $\muF=0$
a point emitter is quenched to $I_{\rm PL}/I_0\approx0.01$, whereas the finite
form factor lifts this to $\approx0.02$ for $R_X=3$~nm and $\approx0.04$ for
$R_X=5$~nm (fixed-dipole benchmark), so the quenching is strong but measurably
size dependent. The
finite-size fingerprint is read most directly from the deviation of the rate from
the point-dipole response evaluated with the same bath [Fig.~\ref{fig:size}(d)]. For $R_X=3$~nm this deviation
is $\approx48\%$ at $z=5$~nm and $\approx18\%$ at $z=10$~nm, and for $R_X=5$~nm it
is $\approx73\%$ and $\approx38\%$ at the same distances. A spacer-thickness
lifetime series resolves this variation and is fit for $R_X$ with the full
Gaussian form factor. Equation~\eqref{eq:leadingR} alone suffices only at the
larger-$z$ end ($z/R_X\gtrsim4$--$5$), so the $R_X=5$~nm case here
($z/R_X=1$--$2$) lies in the crossover and requires the full kernel. (These
illustrative numbers are isotropic; the anisotropic hBN spacer replaces $z$ by
$z_{\rm eff}=\gamma z$.)

Raising the gate to $|\muF|=0.9$~eV, past the threshold
$\hbar\omega/2=0.75$~eV, suppresses the interband-dominated full-RPA ET rate.
Evaluated with the $R_X=3$~nm form factor (not shown) this suppression reaches
$\approx99.5\%$ at $z=5$~nm, lifting $I_{\rm PL}/I_0$ from $\approx0.02$ to
$\approx0.8$. The
point-source gate sweep of Fig.~\ref{fig:rpa}(c) recovers somewhat less
($\approx97\%$ suppression at $z=5$~nm), its higher-$q$ weight being harder to
Pauli-block. Both are $T=0$ estimates. Thermal smearing at $300$~K would reduce
the ET suppression and hence the PL recovery. The recovery is moreover regime
dependent, incomplete in the
strongest-quenching region ($z\lesssim4$~nm, $x\gtrsim80$) but near-complete at
intermediate spacings ($z\sim10$--$15$~nm, $x\sim1$--$5$). All three
signatures---strong quenching, its size dependence, and gate recovery---are
experimentally testable, though the required $|\muF|\sim0.75$--$0.9$~eV
($n\sim4$--$6\times10^{13}~\mathrm{cm}^{-2}$) is most accessible with ionic or
dual-gate techniques or for the lower-energy interlayer excitons
($|\muF|\sim0.55$--$0.7$~eV).

\subsection{Beyond the minimal model: full RPA loss function}
\label{sec:rpa}

To test the minimal model we return to the general overlap formula
Eq.~\eqref{eq:overlap} and evaluate the graphene density response $\chi_G$ within
the RPA, replacing its small-$q$ neutral form $-\mathrm{Im}\,\chi_G\propto
q^2/\omega$ [Eq.~\eqref{eq:minimal}] by the full loss function of doped graphene,
as also used for moir\'e-exciton quenching in Ref.~\cite{Hichri2021}. The
non-interacting polarizability $\Pi(q,\omega)$ at $T=0$ is known in closed
form~\cite{Wunsch2006,Hwang2007,DasSarma2011}. We use their convention, in which
$\Pi(q\!\to\!0,0)=D_0=g E_F/2\pi\hbar^2\vF^2>0$ ($g=4$, with $E_F=|\muF|$ the
Fermi energy and $k_F=E_F/\hbar\vF$) and the
RPA dielectric function is
$\varepsilon(q,\omega)=1+v_q\Pi$ with $v_q=2\pi e^2/(\kappa q)$. Within this
graphene RPA screening function the surrounding dielectric---here the hBN
encapsulation---enters through $\kappa$ in $v_q$ (the average permittivity of the
media on either side of the sheet)~\cite{DasSarma2011}, while the bare bubble
$\Pi$ is a property of the graphene alone; the separate effect of the environment
on propagation from the emitter to graphene is contained in $T(q,\omega)$ and in
the spacer-anisotropy treatment of Sec.~\ref{sec:aniso}. The screened
density response is $\chi_{\rm RPA}=-\Pi/\varepsilon$ and the loss function is
\begin{equation}
  L(q,\omega,\muF)=-\mathrm{Im}\,\chi_{\rm RPA}
  =\mathrm{Im}\!\left[\frac{\Pi(q,\omega)}{\varepsilon(q,\omega)}\right],
  \label{eq:lossrpa}
\end{equation}
which contains the intraband and interband single-particle continua, their
Pauli-blocked phase space at finite $\muF$, and the screened plasmon.

Since the emitter couples through its \emph{bare} near-field potential $\phi_X$
[Eq.~\eqref{eq:overlap}], the relevant kernel is the screened density response
$-\mathrm{Im}\,\chi_{\rm RPA}=\mathrm{Im}[\Pi/\varepsilon]$, not the
energy-loss function $-\mathrm{Im}(1/\varepsilon)$. The two are not independent.
Since $1/\varepsilon=1+v_q\chi_{\rm RPA}$, one has
$-\mathrm{Im}(1/\varepsilon)=v_q\,(-\mathrm{Im}\,\chi_{\rm RPA})$, so an equivalent
formulation using $-\mathrm{Im}(1/\varepsilon)$ would move this factor of
$v_q\propto1/q$ from the response kernel into the source coupling. The transfer rate is then
\begin{equation}
  \GET\propto|D_X(\lX)|^2\int_0^\infty\! q\,dq\,e^{-2qz}\,|F_X(q;\lX)|^2\,L(q,\omega,\muF),
  \label{eq:rate_rpa}
\end{equation}
with $|F_X|^2$ as in Eq.~\eqref{eq:formfactor}.

We verified our implementation of
$\Pi$ against the closed-form static limit~\cite{Ando2006}, $\Pi(q,0)=D_0$ for
$q<2k_F$, the positivity of
$\mathrm{Im}\,\Pi$ in this convention (so that $-\mathrm{Im}\,\chi_{\rm RPA}>0$ is
the energy loss), the neutral-graphene interband form
$\mathrm{Im}\,\Pi=q^2/(4\sqrt{\omega^2-\vF^2q^2})$, and the existence of a
$\sqrt{q}$ plasmon. The polarizability is evaluated at $T=0$. The rate integrals
use $\eta=0$, with the integrable square-root singularities at the kinematic edges
handled by adaptive quadrature with breakpoints there and a momentum cutoff where
$e^{-2qz}$ is negligible. Tightening the adaptive-quadrature tolerance and
extending the momentum cutoff leave the rate integrals and the qualitative scale
separation unchanged. The RPA/minimal comparison is of
the normalized distance shape (relative to $z=20$~nm), not the absolute rate. The
polarizability is the linear-Dirac-cone RPA. At the highest Fermi levels considered ($|\muF|\sim0.9$~eV) its assumptions---a
linear dispersion, electron--hole symmetry, and constant $\vF$---are least
accurate, so lattice-band (trigonal-warping) corrections~\cite{CastroNeto2009} and
the measured optical conductivity may be needed for quantitative modeling there, though the qualitative
gate dependence is unchanged.

\begin{figure*}[t]
  \centering
  \includegraphics[width=\textwidth]{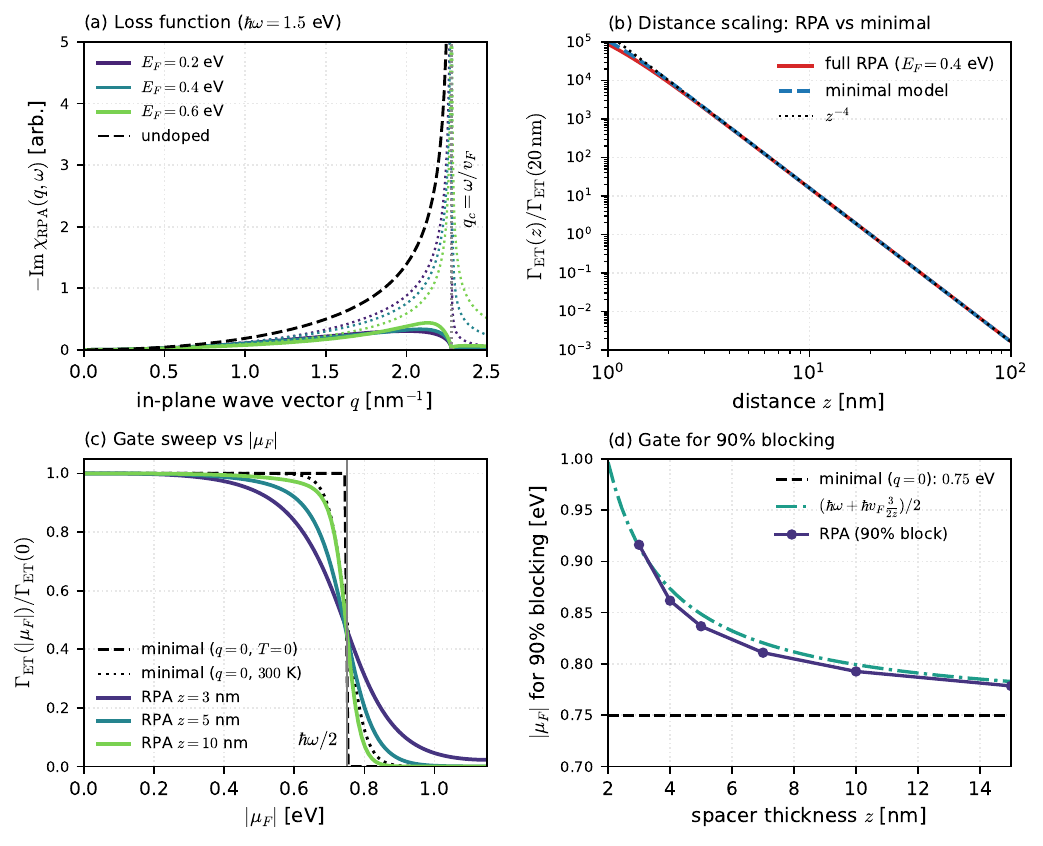}
  \caption{(Color online) Full-RPA validation of the minimal model, for distance and gate
  dependence. (a) $L(q,\omega)$ at $\hbar\omega=1.5$~eV
  for several Fermi levels (solid: screened RPA; dotted: bare bubble; dashed:
  neutral-graphene interband reference). (b) $\GET(z)$ from the RPA loss
  (red) versus the minimal model (blue), each normalized at $z=20$~nm; both
  follow $z^{-4}$ (dotted), deviating only at small $z$. (c) Gate sweep: ET
  suppression $\GET(|\muF|)/\GET(0)$ versus Fermi level from the full RPA at
  $z=3,5,10$~nm (solid), compared with the sharp $q=0$ minimal model (dashed
  $T=0$ step, dotted $300$~K). The RPA suppression is gradual and $z$-dependent,
  with a high-$q$ tail extending past the $q=0$ threshold $\hbar\omega/2=0.75$~eV
  (vertical line). (d) Gate needed for $90\%$ blocking versus spacer thickness:
  the RPA (points) tracks the estimate $(\hbar\omega+\hbar\vF q_{\rm peak})/2$ with
  $q_{\rm peak}=3/2z$ (dash-dotted) and rises well above the $q=0$ value
  ($0.75$~eV, dashed), reaching $\approx0.85$--$0.91$~eV at $z=5$--$3$~nm. Panels
  (c,d) use the point-source form factor ($F_X=1$); a finite exciton size shifts
  the near-field weight to lower $q$, so the gate needed for a given blocking level
  is somewhat lower, making these curves an upper estimate for an extended
  source.}
  \label{fig:rpa}
\end{figure*}

Figure~\ref{fig:rpa}(a) shows the loss function for hBN-encapsulated graphene
($\kappa=4.5$, $\alpha=e^2/\kappa\hbar \vF\approx0.49$). This representative value
agrees to within $\sim15\%$ with the optical
$\kappa_{\rm eff}=\sqrt{\varepsilon_\parallel\varepsilon_\perp}\simeq3.9$ from
Sec.~\ref{sec:aniso}, and the small-$q$ result
below is independent of $\kappa$ since $\varepsilon\to1$ there. At the exciton
emission energy $\hbar\omega=1.5~\mathrm{eV}\gg E_F$, the response is
interband-dominated.
At small $q$---the only momenta that survive the $e^{-2qz}$ near-field filter at
experimentally relevant $z$---the screened loss function converges to the
neutral-graphene interband result, because $\mathrm{Re}\,\Pi\to0$ there and
$\varepsilon\to1$. Doping (Pauli blocking of part of the interband phase space)
and screening only suppress $L$ at intermediate $q$. Consequently the full-RPA
rate follows the $z^{-4}$ law at large $z$ [Fig.~\ref{fig:rpa}(b), fitted slope
$-3.998$] and, for the representative parameters there, its normalized distance
dependence agrees with the minimal model to within a few percent for
$z\gtrsim3$~nm, deviating by $\lesssim10\%$ only at $z\approx2$~nm. The minimal
model Eq.~\eqref{eq:minimal} is therefore quantitatively reliable for the distance
scaling in the regime of interest, and genuine Fermi-level control of the rate at this emission energy
requires reaching the Pauli threshold $|\muF|>\hbar\omega/2$, consistent with
Sec.~\ref{sec:results} above.

The RPA also validates the gate dependence
[Fig.~\ref{fig:rpa}(c,d)]. The suppression with $|\muF|$ is gradual and
$z$-dependent, and near-complete blocking of the near-field momenta
$q\sim q_{\rm peak}=3/2z$ requires $|\muF|$ rising from $\approx0.75$~eV at large
$z$ to $\approx0.85$--$0.91$~eV at $z=5$--$3$~nm. This tracks the analytic estimate
$(\hbar\omega+\hbar\vF q_{\rm peak})/2$ and confirms that the sharp $q=0$ model
underestimates the gate needed at short range.

\subsection{Anisotropic hBN spacer}
\label{sec:aniso}

The dielectric environment is known to influence moir\'e-exciton energy transfer
to graphene~\cite{Hichri2021}. Here we make the role of the spacer anisotropy
explicit. The hBN spacer is a uniaxial dielectric with in-plane permittivity
$\varepsilon_\parallel$ and out-of-plane $\varepsilon_\perp$. Solving the
anisotropic Laplace equation $\varepsilon_\parallel q^2\phi=\varepsilon_\perp
\phi''$ for an in-plane mode $q$ shows that the near field decays vertically as
$e^{-\gamma q z}$ with
\begin{equation}
  \gamma=\sqrt{\varepsilon_\parallel/\varepsilon_\perp},
  \label{eq:gamma}
\end{equation}
so, for the in-plane optical transition dipole relevant to TMDC excitons,
anisotropy enters the rate only through the replacement $z\to\gamma z$ in the
near-field factor (a perpendicular dipole would acquire an additional $\gamma$
prefactor from the vertical derivative, but the in-plane form factor is unchanged),
while the coupling is
screened by $\kappa_{\rm eff}=\sqrt{\varepsilon_\parallel\varepsilon_\perp}$,
taken in the local-dielectric limit where $\kappa_{\rm eff}$ is $q$-independent.

For bulk hBN the relevant response at the optical transition energy
$\hbar\omega=1.5$~eV is the electronic (high-frequency) permittivity,
$\varepsilon_\parallel\simeq4.98$ and $\varepsilon_\perp\simeq3.03$~\cite{Laturia2018},
giving $\gamma\simeq1.28$ and $\kappa_{\rm eff}\simeq3.9$. The static values
$\varepsilon_\parallel^{0}\simeq6.93$ and $\varepsilon_\perp^{0}\simeq3.76$ include
the ionic contribution and apply only below the hBN phonon bands, not at the
exciton energy. As shown in Fig.~\ref{fig:validity}(a,b), the
$z^{-4}$ rate is then suppressed by $\gamma^4\simeq2.7$ relative to an isotropic
treatment, and the graphene quenching distance shrinks by $1/\gamma\simeq0.78$
(a $\sim28\%$ overestimate of the range if anisotropy is neglected).

The anisotropy also propagates into the size extraction. With $z\to\gamma z$,
Eq.~\eqref{eq:leadingR} becomes
$\GET\propto(\gamma z)^{-4}[1-\tfrac52 R_X^2/(\gamma z)^2+\cdots]$ and the
half-suppression sits at $\gamma z\simeq1.6\,R_X$, so a fit using the physical
thickness $z$ in place of the effective thickness $z_{\rm eff}=\gamma z$
underestimates $R_X$ by $1/\gamma$ ($\sim22\%$ for bulk hBN). The effective
thickness $z_{\rm eff}=\gamma z$ should therefore be used when applying
Eq.~\eqref{eq:leadingR} and the crossover of Fig.~\ref{fig:size}(d) to extract
$R_X$ from a measured spacer-thickness series. When $d_0$ is itself calibrated in
the same anisotropic environment, the overall $\gamma^{-4}$ suppression of the
point-dipole rate is absorbed into $d_0$ and must not be applied again; only the
finite-size crossover, set by $F_X(q)$ at $q\sim1/\gamma z$, then carries the
effective thickness.

\subsection{Plasmon-resonant regime and validity boundary}
\label{sec:plasmon}

\begin{figure*}[t]
  \centering
  \includegraphics[width=\textwidth]{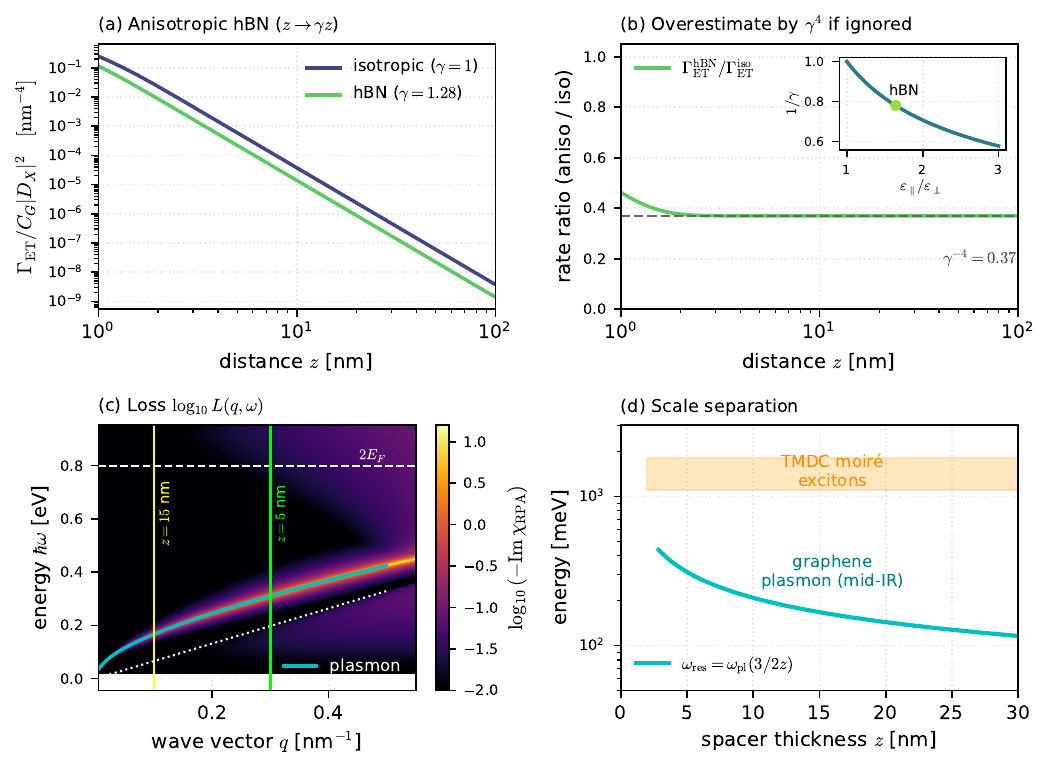}
  \caption{(Color online) Validity checks of the interband minimal model. Top row (anisotropic
  hBN spacer): (a) point-dipole $\GET(z)$ for an isotropic spacer ($\gamma=1$)
  and bulk hBN ($\gamma=1.28$; anisotropy acts as $z\to\gamma z$); (b) the rate
  ratio plateaus at $\gamma^{-4}\simeq0.37$ at large $z$, inset the
  quenching-distance factor $1/\gamma$ versus
  $\varepsilon_\parallel/\varepsilon_\perp$ with bulk hBN marked. Bottom row
  (plasmon regime): (c) RPA loss $\log_{10}L(q,\omega)$ at $E_F=0.4$~eV showing
  the $\sqrt{q}$ plasmon ridge below the interband region, with the representative
  near-field momentum $q\simeq3/2z$ (the rate-integrand peak) for $z=5,15$~nm
  (white dotted line: $\omega=\vF q$; white
  dashed: $2E_F$); (d) the resonant energy
  $\omega_{\rm pl}(3/2z)$ versus spacer thickness lies in the mid-infrared, well
  separated from the TMDC moir\'e-exciton band.}
  \label{fig:validity}
\end{figure*}

Besides the interband continuum, the RPA loss function~\eqref{eq:lossrpa} hosts a
$\sqrt{q}$ plasmon pole in the single-particle gap [Fig.~\ref{fig:validity}(c),
rendered with a small broadening $\eta=0.05\,E_F$],
the basic collective mode of graphene plasmonics~\cite{Grigorenko2012,GarciaDeAbajo2014}.
Energy transfer is resonantly enhanced---and deviates from the $z^{-4}$ law---when
the emission energy matches the plasmon at the momenta admitted by the near-field
filter $e^{-2qz}$. That filter has exponential cutoff scale $q\sim1/2z$, whereas
the rate integrand $q^3e^{-2qz}$ peaks at the larger $q_{\rm peak}=3/2z$ (with mean
$\langle q\rangle=2/z$). We take this integrand peak as the representative
near-field momentum, so that $\hbar\omega\simeq\hbar\omega_{\rm pl}(q_{\rm peak})$.
These momentum scales ($1/2z$, $1/z$, $3/2z$) differ only by numerical factors of
order unity and all track the inverse spacer thickness. Graphene plasmons are,
however, mid-infrared. At $E_F=0.4$~eV the resonant energy
$\omega_{\rm res}(z)=\omega_{\rm pl}(3/2z)$ ranges only $\sim140$--$310$~meV for
$z=20$--$5$~nm [Fig.~\ref{fig:validity}(d)], still well below TMDC exciton energies
($1.1$--$1.8$~eV). The plasmon channel is therefore
off-resonant for moir\'e excitons---which lie well within the interband regime
where the minimal model holds---and would dominate only for infrared emitters or
much higher doping. This scale separation delimits the validity of the
interband-based theory. For visible/near-infrared moir\'e-exciton PL, the
collective mode can be neglected and Eq.~\eqref{eq:minimal} applies. Coupling to
the plasmon becomes a distinct, distance-tunable resonance only in the infrared.

Throughout we use the non-retarded (electrostatic) near field and the
correspondingly non-retarded $\sqrt{q}$ plasmon dispersion. This is justified
because, for the nanometric strong-quenching regime emphasized here, the
representative near-field momentum $q_{\rm peak}=3/2z$ lies well above the light
line $\omega/c$, where retardation would modify the dispersion. Retardation
becomes important only at much longer wavelengths, where it drives a crossover
from linear to sublinear plasmon dispersion in graphene~\cite{Deng2015}. In the
mid-infrared the hBN dielectric response is itself strongly frequency-dependent,
and the graphene plasmon can hybridize with hBN
phonon-polaritons~\cite{Brar2014}.
Figure~\ref{fig:validity}(d) is therefore a bare-graphene order-of-magnitude
estimate, which does not affect the visible/near-infrared conclusion.

\section{Discussion}
\label{sec:discussion}

The theory above is deliberately minimal, and a few assumptions delimit its scope.
First, the Gaussian form factor~\eqref{eq:formfactor} is the simplest
localized-exciton model: a realistic moir\'e wave function alters the detailed
crossover through higher moments of $F_X(q)$, but the leading long-distance
deviation---and hence the extracted $R_X$---is unchanged. Second, the coherent
scaling $|D_X|^2\propto\lX^2$ (the giant oscillator strength) is an idealization.
It holds only in weak confinement, where the center-of-mass envelope $\lX$ exceeds
the internal exciton radius $a_X\sim1$--$2$~nm (the electron--hole
relative-coordinate size); the smallest $\lX$ plotted in Fig.~\ref{fig:size} fall
below this bound and are included only to illustrate the size dependence, not as
physical cases. A coherent exciton radiates like a phased antenna array: the unit-cell transition
dipoles add \emph{in phase}, so the bright-state $|D_X|^2$ grows with the coherent
area. This in-phase addition breaks down beyond the phase-coherence length
$L_\phi$, the moir\'e-domain size $L_{\rm domain}$, or the optical wavelength, so
$|D_X|^2$ saturates at $\min(\lX^2,L_\phi^2,L_{\rm domain}^2,A_{\rm opt})$ rather
than growing without limit---equivalently, the oscillator-strength sum rule
conserves the total and only redistributes it. Real excitons are only
partially coherent---phonons and disorder degrade the coherence---so $|D_X|^2$
grows with $\lX$ more weakly than the ideal $\lX^2$ law. Their absolute size
dependence therefore lies between the coherent and fixed-dipole curves of
Fig.~\ref{fig:size}.

The bath and the emitter contribute in distinct distance regimes.
The intermediate-distance $z^{-4}$ asymptote reflects the small-$q$ response of the
graphene bath, independent of the emitter's internal structure. It is therefore
recovered for any localized emitter---a molecule or an exciton wave packet---once
$z$ exceeds its coherence length, within the nonretarded window
$R_X,\qc^{-1}\ll z\ll c/\omega\simeq130$~nm at $\hbar\omega=1.5$~eV. By contrast, an ideal free exciton---translationally invariant, with a sharp
center-of-mass momentum $Q$ so that $P_X(q)\propto\delta(q-Q)$---is a distinct
limit. Its rate samples the bath at that single momentum,
$\propto e^{-2Qz}[-\mathrm{Im}\,\chi_G(Q,\omega)]$, and vanishes as $Q^2$ for a
bright ($Q\to0$) state, since $-\mathrm{Im}\,\chi_G\propto q^2$. A delocalized
bright exciton therefore barely couples to graphene: it is the finite-$q$ spread
of a \emph{localized} wave packet that provides the coupling. The finite-size
deviation and its $R_X$ dependence are thus where the tunable moir\'e confinement
is essential.

The same formalism applies to disorder-localized monolayer excitons (with $\lX$ replaced
by a random $\ell_{\rm dis}$). The advantage of moir\'e excitons is that $\lX$ is
set by twist angle and moir\'e depth rather than by uncontrolled disorder. In
practice both contributions coexist, and the resulting spatial inhomogeneity of
moir\'e-exciton PL can be mapped and statistically disentangled by hyperspectral
imaging and descriptor-correlation analyses~\cite{Ahmad2026,Wakabayashi2026desc}. The
theory should thus be read as a general exciton-to-graphene energy-transfer
formalism, with moir\'e excitons providing the most controlled realization of a
finite and tunable emitter form factor.

The same reasoning indicates what changes if the graphene gate is replaced by
multilayer graphene or graphite, as in many real devices. The $z^{-4}$ law is
specific to a single 2D sheet. An independent-layer sum over the interlayer spacing
$d\simeq0.34$~nm gives
$\sum_{n\ge0}\GET^{(1)}(z+nd)\sim\int_z^\infty dz'/z'^4\sim1/(3d\,z^3)$ for
$z\gg d$, i.e. a crossover toward $\GET\propto z^{-3}$---the surface-energy-transfer
law for a molecule near an absorbing half-space~\cite{Chance1978}. Because
$z^{-3}$ falls off more slowly than $z^{-4}$, graphite quenches more strongly and
over a longer range (by a factor $\sim z/3d\sim5$ at $z=5$~nm). The emitter-side sensitivity to
$R_X$ persists, but the softer $\sum_n e^{-2q(z+nd)}\simeq e^{-2qz}/(2qd)$ kernel
lowers the leading correction coefficient from $5/2$ to $\simeq3/2$. Gate
tunability is largely lost, because $c$-axis screening confines doping to the top
one or two layers and Pauli blocking then removes only a fraction $\sim d/z$ of the
response. An electrically switchable bath is thus specific to monolayer (or
few-layer) graphene, whereas graphite is a permanently active, longer-ranged
quencher. This independent-layer picture is a scaling argument rather than a
quantitative prediction, and a realistic treatment would use the graphite bulk and
surface dielectric response.

The sharp Pauli-blocking factor~\eqref{eq:pauli} omits the intraband (Drude) and
plasmonic contributions of the full RPA polarizability. As shown in
Secs.~\ref{sec:rpa} and~\ref{sec:plasmon}, at $\hbar\omega\gg E_F$ these are
confined to intermediate momenta or to the mid-infrared, leaving the minimal model
accurate for visible/near-infrared excitons at $z\gtrsim3$~nm. Finite temperature
and hBN anisotropy [Eqs.~\eqref{eq:thermal} and~\eqref{eq:gamma}] likewise fit
within the overlap form~\eqref{eq:overlap} without altering its structure.

A further channel
not treated here is direct interlayer charge transfer, which can compete with
energy transfer in directly contacted graphene--TMDC
stacks~\cite{Froehlicher2018}. The present theory addresses the nonradiative
\emph{energy}-transfer pathway, which dominates once a few nanometers of hBN
suppress the wave function overlap required for tunneling while leaving the
long-range Coulomb coupling intact. Consistent with this picture, a recent
distance-resolved study of WS$_2$/graphene found F\"orster-type energy transfer
to dominate at finite spacer thickness, with a Dexter (tunneling) contribution
only at direct contact~\cite{Tebbe2024}. Energy transfer that quenches and spectrally
filters the photoluminescence of an atomically thin semiconductor placed next to
graphene has indeed been observed directly~\cite{Lorchat2020}, and rapid
interlayer energy transfer between dissimilar TMDC monolayers is likewise well
established experimentally~\cite{Kozawa2016} and theoretically~\cite{Manolatou2016}.

Two features make the predictions experimentally testable. First, because the
relative PL intensity and lifetime are predicted to track each other
[Eq.~\eqref{eq:pl}], a measured correlation between brightness and lifetime as a
function of spacer thickness is a diagnostic signature of ET rather
than of varying disorder, especially when combined with gate tuning. In practice
$I_{\rm PL}$ is also affected by absorption of the excitation light, collection
efficiency, exciton diffusion, and trap capture, so the
equality $I_{\rm PL}/I_0=\tau_{\rm PL}/\tau_0$ should be read as a diagnostic
limit in which only $\GET$ varies. Lifetime measurements are insensitive to
excitation absorption and collection efficiency and therefore provide the cleaner
test, although diffusion, trap capture, and state feeding can still modify the
decay dynamics and must be controlled. Second, the gate dependence provides an \emph{in situ} control. The
same device can be driven between the active and Pauli-blocked regimes, and the
progressive PL recovery across this transition isolates the
graphene contribution. The required $|\muF|\sim0.6$--$0.9$~eV, depending on
emission energy and spacer thickness, is demanding for conventional hBN-gated
devices at visible exciton energies, but becomes more accessible for the
lower-energy interlayer excitons ($\hbar\omega\sim1.1$--$1.4$~eV,
$|\muF|\sim0.55$--$0.7$~eV), for infrared emitters, or for devices using strong
electrostatic or ionic gating. Gate control of interlayer-exciton dynamics in
such devices is already established~\cite{Jauregui2019}.
The universal distance scaling and gate control already demonstrated for point
emitters~\cite{Gaudreau2013,Tielrooij2015} support the underlying mechanism.
The new, testable element here is the dependence on the moir\'e-exciton
transition-polarization radius $R_X$.

A concrete protocol makes the extraction well posed and circumvents the
phenomenological normalization. Because the graphene gate strongly modulates the
nonradiative channel between active and approximately Pauli-blocked states, one
can measure the time-resolved PL of the
\emph{same spectroscopically identified exciton family} at each spacer thickness $z$ in both the active
($\muF=0$) and Pauli-blocked ($|\muF|>\hbar\omega/2$) states and form the
difference
\begin{equation}
  \Delta\GET(z)\equiv\frac{1}{\tau_{\rm active}(z)}-\frac{1}{\tau_{\rm blocked}(z)}.
  \label{eq:extract}
\end{equation}
Only $\GET$ responds to the gate; the radiative rate, the intrinsic nonradiative
rate, and any disorder and diffusion losses are gate-independent, hence identical
in the active and blocked states, and cancel in the difference---leaving the
change in $\GET$. The subtraction is not exact. Because the interband threshold is $q$-dependent,
the blocked state keeps a small finite-$q$ interband tail, so $\Delta\GET$ equals
$\GET$ only up to this residual,
$\Delta\GET(z)\propto|D_X|^2\int q\,dq\,e^{-2qz}|F_X(q)|^2[L(q,\omega,\mu_{\rm active})-L(q,\omega,\mu_{\rm blocked})]$,
most pronounced at short spacings where $q_{\rm peak}=3/2z$ is largest. The simple
leading-order fit [Eq.~\eqref{eq:leadingR}] is therefore rigorous only for complete
blocking; in general one extracts $R_X$ by fitting against the full-RPA difference
kernel $L(q,\omega,\mu_{\rm active})-L(q,\omega,\mu_{\rm blocked})$
[Sec.~\ref{sec:rpa}, Fig.~\ref{fig:rpa}(c,d)].

The subtraction isolates $\GET$ only if gating the graphene leaves the exciton
itself unchanged; otherwise the active and blocked states differ by more than the
loss channel. Two effects must be avoided: the gate field Stark-shifting the
exciton energy, and the changed graphene screening renormalizing $D_X$, $R_X$, or
$\omega_X$. Both are suppressed by holding the exciton energy, charge state, and
displacement field fixed, ideally through dual-gate compensation. This is checked from the exciton energy,
linewidth, polarization, and radiative spectral weight remaining unchanged. The
selected line should also decay approximately single-exponentially. If it does
not, the active and blocked transients must be fit together with a shared
multilevel rate-equation model, since graphene also changes the lifetimes of the
states that feed the line, not only that of the line itself.

Taking the ratio $\Delta\GET(z)/\Delta\GET(z_{\rm ref})$ to a reference thickness
then removes the unknown oscillator strength $|D_X|^2$, the absolute prefactor
$C_G$, and $d_0$. The remaining normalized distance dependence is fit for
$R_X$ against the known graphene kernel and the effective thickness
$z_{\rm eff}=\gamma z$ [Eq.~\eqref{eq:leadingR} and
Secs.~\ref{sec:size},~\ref{sec:aniso}], not against a bare $z^{-4}$ law.
Here $z$ is the distance from the exciton transition-polarization centroid to the
graphene plane, not the nominal hBN thickness. Restricting the sweep to
$z\gtrsim2$--$3$~nm keeps the measurement in the F\"orster (energy-transfer)
regime, away from the Dexter/charge-transfer channel at direct
contact~\cite{Froehlicher2018,Tebbe2024}. The measurement sensitivity, the
parameter correlations that set the uncertainty on $R_X$, the spacer offset, and
the device-to-device statistics required by a multi-thickness series are collected
in Appendix~\ref{app:protocol}. Being both differential and self-referenced, the scheme extracts $R_X$ from the
deviation of the finite-size signal from the calibrated point-source difference
kernel, without recourse to an absolute rate.

For the same reason, the PL maps of
Fig.~\ref{fig:gatepl}(e,f) should be read as scaling illustrations. They use the
fixed-dipole benchmark and the representative $d_0$. The RPA comparison
validates the graphene loss function and the distance dependence of the kernel,
but not the absolute exciton transition source or prefactor.

\section{Conclusion}
We have formulated nonradiative energy transfer from moir\'e excitons to graphene
as the overlap between the exciton near-field spectrum and the long-wavelength
electronic density response of graphene, weighted by an exciton form factor. The
framework reproduces the known $z^{-4}$ law in the point-dipole limit. Its leading
deviation from that law, measured relative to the calibrated point-dipole
response, is governed at long distance by the transition-polarization radius $R_X$
and thus probes the moir\'e-exciton extent. In the ideal fully coherent limit the
size dependence is non-monotonic, peaking at $\lX\approx z$. Finally, Pauli
blocking makes the nonradiative channel electrically tunable. The cleanest
experimental test is a gate-referenced, spacer-thickness-dependent lifetime
measurement. Its active--blocked difference isolates the graphene-induced change in the decay
rate; the deviation of its normalized distance dependence from the calibrated
point-dipole response then yields $R_X$ at long distance.
With an assumed transition-polarization profile, the full crossover can be used as
well. Gate tuning of the
bath is most favorable for low-energy interlayer excitons and, more generally,
for infrared emitters, where the Pauli threshold $|\muF|\sim\hbar\omega/2$ and
the graphene plasmon are most accessible. In van der Waals exciton devices, a
graphene gate should therefore be regarded not as a passive electrostatic
element but as a tunable 2D nonradiative reservoir whose
long-wavelength response can be probed through exciton photoluminescence
quenching.

\begin{acknowledgments}
This work was supported by JSPS KAKENHI (Grants No. JP25K01609 and No. JP22H05473).
\end{acknowledgments}

\appendix

\section{Neutral-graphene response and the $q^3$ kernel}
\label{app:kernel}

The minimal kernel of Eq.~\eqref{eq:minimal} follows from the standard
small-$q$, Dirac-cone density response of neutral graphene. At $T=0$ and for
$\omega>\vF q$, the imaginary part of the noninteracting polarizability is
\begin{equation}
  -\mathrm{Im}\,\chi_0(q,\omega)
  =\frac{g}{16}\,\frac{q^2}{\sqrt{\omega^2-\vF^2q^2}}\,
  \Theta(\omega-\vF q),
  \label{eq:imchi0}
\end{equation}
with degeneracy $g=4$ (we set $\hbar=1$ in this Appendix). The step
$\Theta(\omega-\vF q)$ is kinematic. A Dirac interband pair of total energy
$\omega$ carrying momentum $q$ requires $|\bm k|+|\bm k+\bm q|=\omega/\vF$, which
by the triangle inequality is $\geq q$, so no such pair---and hence no
absorption---exists above the cutoff $\qc=\omega/\vF$. The same Dirac-cone
polarization, evaluated statically, gives the $q$-linear interband result
$\Pi^0(q)=g\,q/(16\vF)$~\cite{Ando2006}, sharing the $g/16$ prefactor. For
$\omega\gg \vF q$---the regime selected by the near-field filter at
experimentally relevant $z$---this reduces to
$-\mathrm{Im}\,\chi_0\simeq (g/16)\,q^2/\omega\propto q^2/\omega$, the form used
in the main text. This long-wavelength interband response is the same physics as
the frequency-independent universal optical conductance $\sigma_0=\pi e^2/2h$ of
graphene, predicted by Ando \emph{et al.}~\cite{Ando2002} and
measured in Ref.~\cite{Nair2008}, here resolved at finite $q$.

A transition dipole at height $z$ couples to graphene through its bare Coulomb
potential. For a dipole perpendicular to the plane, the in-plane Fourier
component of the potential is $\phi_X(q,z)\propto e^{-qz}$ (the dipole derivative
$\partial_z$ brings a factor $q$ that cancels the $1/q$ of the 2D
Coulomb kernel $2\pi/q$), so $|\phi_X|^2\propto e^{-2qz}$. An in-plane dipole
gives $|\phi_X|^2\propto\cos^2\!\theta\,e^{-2qz}$, whose azimuthal average is
$\tfrac12 e^{-2qz}$. The orientation enters only through this constant, absorbed
into $C_G(\omega)$. Inserting both factors into the overlap form
Eq.~\eqref{eq:overlap} with the planar measure $d^2q=2\pi q\,dq$ gives
\begin{equation}
  \GET\propto\int q\,dq\,\underbrace{e^{-2qz}}_{|\phi_X|^2}\,
  \underbrace{\frac{q^2}{\omega}}_{-\mathrm{Im}\,\chi_0}
  =\frac{1}{\omega}\int dq\,q^3 e^{-2qz},
\end{equation}
i.e. Eq.~\eqref{eq:minimal} with $C_G(\omega)\propto1/\omega$, and
$\int_0^\infty q^3e^{-2qz}dq=3/(8z^4)$ reproduces the Swathi--Sebastian
$z^{-4}$ law~\cite{Swathi2008,Swathi2009}. The chirality factor of the Dirac
matrix elements is contained in Eq.~\eqref{eq:imchi0} and likewise affects only
the prefactor, not the $z$-dependence.

\section{Gaussian exciton form factor}
\label{app:formfactor}

The optical transition polarization of an exciton is linear in its center-of-mass
amplitude. Coarse-grained over the envelope scale,
$\langle 0|\hat{\bm P}(\bm R)|X\rangle\propto\psi_X(\bm R)$, with the microscopic
interband dipole and the relative-coordinate wave function at zero
electron--hole separation setting the local dipole prefactor. The
near-field source is therefore the transition-polarization density
$\bm P_X(\bm R)=\bm u\,P_X(\bm R)$ with $P_X(\bm R)\propto\psi_X(\bm R)$,
whose in-plane Fourier transform is $P_X(\bm q)=D_X\,F_X(q;\lX)$ with the
normalized form factor
\begin{equation}
  F_X(q;\lX)=\frac{\int d^2R\,\psi_X(\bm R)\,e^{-i\bm q\cdot\bm R}}
  {\int d^2R\,\psi_X(\bm R)}.
\end{equation}
The exciton center of mass sits in the moir\'e potential, which is harmonic near a
minimum. Taking the origin at that minimum, $\bm R$ is the center-of-mass
displacement from it and $V\simeq\tfrac12 M\omega_0^2 R^2$ with $R=|\bm R|$; its ground state is the normalized 2D
Gaussian envelope $\psi_X(\bm R)=(\sqrt{\pi}\,\lX)^{-1} e^{-R^2/2\lX^2}$
($\int d^2R\,|\psi_X|^2=1$), with $\lX=\sqrt{\hbar/M\omega_0}$ the oscillator length
set by the moir\'e curvature ($M$ the exciton mass, $\omega_0$ the trap frequency).
This gives $F_X(q;\lX)=e^{-q^2\lX^2/2}$, so the rate carries
$|F_X(q;\lX)|^2=e^{-q^2\lX^2}$, Eq.~\eqref{eq:formfactor}. The total transition dipole is
$D_X=\int d^2R\,P_X(\bm R)\propto\int d^2R\,\psi_X(\bm R)\propto\lX$
(all elements share the direction $\bm u$), i.e. $|D_X|^2\propto\lX^2$. A
coherent exciton delocalized over the area $\lX^2$ carries a giant oscillator
strength.

A different envelope changes $F_X(q;\lX)$
only at $q\gtrsim1/\lX$ and does not alter the conclusion that the distance
dependence probes the finite exciton size, quantified in general by the radius
$R_X$ of Sec.~\ref{sec:lowq}.

For comparison, the fixed-total-dipole benchmark treats the emitter as a rigid
dipole $D_X=\mathrm{const}$ spread with a normalized Gaussian
transition-polarization profile $P_X(\bm R)\propto e^{-R^2/\lX^2}$, whose width
coincides numerically with that of $|\psi_X(\bm R)|^2$ for this Gaussian $\psi_X$
but is not interpreted as a probability density. Its Fourier transform is
$F_X(q;\lX)=e^{-q^2\lX^2/4}$, so $|F_X(q;\lX)|^2=e^{-q^2\lX^2/2}$.
This is a phenomenological rigid-dipole distribution rather than a coherent
quantum transition polarization. It mimics the loss of coherent-area enhancement
by holding the total dipole fixed, and isolates pure momentum filtering without the
$\lX^2$ prefactor. It recovers the point-dipole $z^{-4}$ law for every $\lX$ at
large $z$.

\section{Practical aspects of the extraction protocol}
\label{app:protocol}

The differential extraction protocol of Sec.~\ref{sec:discussion} is sensitivity-
and statistics-limited in ways that set the achievable uncertainty on $R_X$.

\emph{Sensitivity and parameter correlations.} The extraction is
sensitivity-limited at large $z$. Since
$|\partial\ln\GET/\partial R_X|\simeq5R_X/z^2$ is $0.15$, $0.067$, and
$0.038~\mathrm{nm}^{-1}$ at $z=10$, $15$, and $20$~nm for $R_X=3$~nm, resolving
$R_X$ to $\sim0.5$~nm near $z=15$~nm requires measuring the rate to a few percent,
exactly where $\GET$ is smallest. Moreover $R_X$ correlates with the offset $z_0$, the
anisotropy $\gamma$, $D_X$, the fourth moment $\langle R^4\rangle_P$, and the
blocked-state residual. Turning these correlations into a realistic uncertainty on
$R_X$ then requires a joint (Fisher or profile-likelihood) fit of a
multi-thickness series spanning both the crossover ($z\sim R_X$) and the
asymptotic ($z\gg R_X$) regions.

\emph{Spacer offset.} The distance $z$ is measured from the exciton
transition-polarization centroid. At nanometre spacings the finite
TMDC-layer thickness, the interlayer-exciton position, and monolayer steps in the
spacer all shift it and enter the $z^{-4}$ scaling. The physical distance is
$z_{\rm eff}=\gamma(z_{\rm hBN}+z_0)$ with an offset $z_0$, and since
$\delta\GET/\GET\simeq-4\,\delta z/z$, a $0.2$~nm error at $z=3$~nm shifts the
rate by $\sim27\%$, comparable to the finite-size signal. The offset $z_0$ should
therefore be fit as a nuisance parameter or calibrated independently, since it
correlates strongly with $R_X$.

\emph{Device-to-device statistics.} The ratio removes $D_X$ only if the same
exciton state and oscillator strength are maintained across the spacer-thickness
series. Because a thickness series generally means different devices---with
possible variation in twist angle, registry, strain, and disorder---the condition
is best approached with a laterally stepped hBN spacer on a single heterobilayer,
or by calibrating the oscillator strength for each device independently. The gate
subtraction removes the intra-device baseline but not this device-to-device
variation of $D_X$. For a laterally stepped spacer, exciton diffusion can also mix
regions of different $z$ if the diffusion length approaches the terrace width or
the spot size, because strongly quenched regions act as sinks. Each terrace should
therefore exceed the diffusion length, and the measurement be taken away from step
edges and at low excitation density.

Because a given localized site cannot be
placed at several spacer thicknesses, the comparison is across statistically
matched lines, with $R_{X,i}=R_X+\delta R_i$ and $D_{X,i}=D_X+\delta D_i$ varying
from line to line. Measuring several lines per thickness and fitting them as a
random-effects ensemble separates the mean $R_X$ from this scatter.

\bibliographystyle{apsrev4-2}
\bibliography{refs}

\end{document}